\documentclass[12pt,epsfig]{article}
\usepackage{graphicx}

\usepackage[latin1]{inputenc}
\usepackage{amsmath}
\usepackage{amsfonts}
\usepackage{amssymb}
\usepackage{graphicx}
\DeclareGraphicsExtensions{.pdf,.png,.jpg}

\usepackage{epstopdf}
\def\beq{\begin{equation}}
\def\eeq{\end{equation}}
\def\bea{\begin{eqnarray}}
\def\eea{\end{eqnarray}}

\begin{document}

\begin{center}
  {\Large \bf Role of $PT$-symmetry in understanding Hartman effect }
\vspace{1.3cm}

{\sf   Mohammad Hasan  \footnote{e-mail address: \ \ mhasan@isro.gov.in, \ \ mohammadhasan786@gmail.com}$^{,3}$,
Vibhav Narayan Singh \footnote{e-mail address: vibhav.ecc123@gmail.com}
Bhabani Prasad Mandal \footnote{e-mail address:
\ \ bhabani.mandal@gmail.com, \ \ bhabani@bhu.ac.in  }}

\bigskip

{\em $^{1}$Indian Space Research Organisation,
Bangalore-560094, INDIA \\
$^{2,3}$Department of Physics,
Banaras Hindu University,
Varanasi-221005, INDIA. \\ }

\bigskip
\bigskip

\noindent {\bf Abstract}

\end{center}

The celebrated Hartman effect, according to which, the tunneling time through a opaque barrier is independent of the width of the barrier for a sufficiently thick barrier, is not well understood theoretically and experimentally till today. In this work we attempt to through some light to understand the mystery behind this paradoxical result of tunneling .For this purpose  we calculate the tunneling time from a layered non-Hermitian system to examine the effect of $PT$-symmetry over tunneling time. We explicitly find that for system respecting $PT$-symmetry, the tunneling time saturates with the thickness of the $PT$-symmetric barrier and thus shows the existence of Hartman effect. For non PT-symmetric case, the tunneling time depends upon the thickness of the barrier and Hartman effect is lost. We further consider the limiting case in which the non-Hermitian system reduces to the real barrier to show  that the Hartman effect from a real barrier  is due to $PT$-symmetry (of the corresponding non-Hermitian system) .

\medskip
\vspace{1in}
\newpage

\section{Introduction}        
The study of non-Hermitian system in quantum mechanics started as a mathematical curiosity. In the year 1998, it was shown that a non-Hermitian system which respect $PT$-symmetry can yield  real energy eigen values \cite{ben4}. It was also found that a fully consistent quantum theory can be developed for non-Hermitian system in a modified Hilbert space through the restoration of equivalent Hermiticity and the unitrary time evolution \cite{mos, benr}. These theoretical works towards the consistency of non-Hermitian quantum mechanics (NHQM) strongly paved the way forward for NHQM to be the topic of frontier research in different areas in the last two decades \cite{nh1}-\cite{nh7}. Due to the analogy of the Schrodinger equation with certain wave equation in optics, the phenomena of NHQM can also be mapped to the analogous phenomena in optics. This lead to the possibility of experimental observation of the theoretical predictions of NHQM.  This has been indeed the case and some of the predictions of NHQM have been observed in optics   \cite{ opt1}-\cite{kotto}. The  realizations of NHQM phenomena  have ignited huge interest to study the subject both theoretically and experimentally. The study of non-Hermitian system in optics has become a constant theme of further research and advancement in the subject.

The advancement in NHQM is one of the most recent developments in quantum mechanics. However one of the  earliest studied problems of quantum mechanics, the quantum tunneling \cite{ nordheim1928, gurney1928, condon, wigner_1955, david_bohm_1951}, suffers with a paradox till today. How much time does a particle take to tunnel through a classically forbidden potential is still an open problem both theoretically and experimentally. In the year 1962, Hartman studied the problem of tunneling time by using stationary phase method (SPM) for metal-insulator-metal sandwich and showed that the tunneling time for opaque barrier is independent of the thickness for sufficiently thick barrier \cite{hartman_paper}. This is known as Hartman effect i.e. the saturation of tunneling time for an opaque barrier with the barrier thickness. Soon, this was also confirmed by an independent study by Fletcher \cite{fletcher}. Due to this paradox, various different authors proposed  new definitions of tunneling time to account for the inconsistency (see \cite{hg_winful} and references therein). However, so far no satisfactory definition of tunneling time has been found  that agrees with the experimental results. 

The calculation of tunneling time by the method of SPM for multi-barrier real potential shows that tunneling time is independent of the inter-barrier separation in the limit of large thickness of the barrier \cite{generalized_hartman, esposito_multi_barrier}.  This is called as generalized Hartman effect in which the tunneling time is also independent of the inter-barrier separation for the tunneling through sufficiently thick opaque multi-barrier. For critical comments on generalized Hartman effect, see  \cite{questions_ghf1,questions_ghf2,questions_ghf3}. Various attempts have been made to test the finding of the theoretical results of the tunneling time.  Initial experiments have indicated the superluminal nature of the tunneling time and found to be insensitive to the thickness of the tunneling region \cite{sl_prl, nimtz, ph, ragni, sattari, longhi1, olindo}. This superluminal nature of the tunneling time is not at variance with Special Relativity and the phenomena of this kind have been discussed in a number of papers ( see \cite{barbero2000,recami2000} and references therein).  The tunneling time found to be paradoxically short for the case of double barrier optical grating \cite{longhi1} and double barrier photonic band gap \cite{longhi2}.   
The reason for Hartman effect is not clear to the present day. A reshaping of the incident wave as it interact with the barrier has been proposed as a possible reason for the occurrence of Hartman effect \cite {reshaping}. Also, Hartman effect doesn't occur in space fractional quantum mechanics \cite {tt_sfqm_1,tt_sfqm_2}. 

To the best of our knowledge, the method of SPM has always shown the existence of Hartman effect from a real barrier potential (single  barrier or multi-barrier). However, for a complex, non-PT symmetric barrier potential of the form $V_{1}+ i V_{2}$, it has been shown that Hartman effect doesn't occur and the tunneling time depends upon the barrier thickness \cite{complex_barrier_tunneling} .   Also it is shown in \cite{hartman_layered}, when the complex potential is in the form of a layered $PT$-symmetric potential, the Hartman effect does occur for single as well for periodic multi-barrier systems.  The result of \cite{complex_barrier_tunneling}, \cite{hartman_layered} and the Hartman effect from real barrier have motivated us to study the role of $PT$-symmetry in the occurrence of Hartman effect.  We study the Hartman effect from a layered $PT$-symmetric potential and show that the occurrence of Hartman effect from a real barrier can be understood as the special limiting case of Hartman effect from a $PT$-symmetric complex system . We also study the tunneling time from a non PT-symmetric potential at the symmetry breaking threshold and show that Hartman effect doesn't occur when $PT$-symmetry is broken. Hartman effect is restored when $PT$-symmetry is respected .  These results give strong indication that $PT$-symmetry  plays an important role for the occurrence of Hartman effect.  We  further have  shown explicitly that PT symmetry is crucial even for the real barrier which is shown to be the special limiting case of a layered $PT$-symmetric  complex system. 

\paragraph{}
We organize our paper as follows: In section \ref{spm}, we introduce the reader about stationary phase method of calculating the tunneling time. In section \ref{hf_pt_discussions} we discuss the Hartman effect from a `unit' PT-symmetric system and a layered $PT$-symmetric system made by the periodic repetitions of the `unit' $PT$-symmetric system. In subsection \ref{real_barrier_section}, we show that the Hartman effect from real barrier is the special limiting case of our layered $PT$-symmetric system. In section \ref{non_pt_section} , we calculate the tunneling time from a non $PT$-symmetric system at the $PT$-symmetry breaking threshold to show that when $PT$-symmetry is broken, Hartman effect is lost. We discuss the results in   \ref{results_discussions} .  Detail mathematical  steps in obtaining various results are provided in Appendix.

\section{Tunneling time and Hartman effect}
\label{spm}
This section briefly introduces the reader about the stationary phase  method (SPM) to calculate the  tunneling time \cite{dutta_roy_book}. In SPM, the tunneling time is defined as the time difference between the peak of the incoming and outgoing wave packet as the wave packet traverse through the potential barrier.  To understand this quantitatively,  consider a normalized Gaussian wave packet $G_{k_{0}} (k)$ of mean momentum $\hbar k_{0}$.  For $t>0$, the wave packet is given by,
\begin{equation}
\int G_{k_{0}} (k)e^{i(kx-\frac{Et}{\hbar})}dk.
\label{localized_wave_packet}
\end{equation}
In the above $k=\sqrt{2mE}$. The wave packet is propagating to positive $x$-direction and interact with the potential barrier $V(x)$ ($V(x)=V $ for $0 \leq x \leq b$ and zero elsewhere). The transmitted wave packet is given by,
\begin{equation}
\int G_{k_{0}} (k) \vert A(k) \vert e^{i(kx-\frac{Et}{\hbar} +\theta(k))}dk .
\label{emerged_wave_packet}
\end{equation}
Where $A(k)=\vert A(k) \vert e^{i\theta (k)}$ is the transmission coefficient for the potential barrier $V(x)$. By the method of SPM, the tunneling time $\tau$ is given by
\begin{equation}
\frac{d}{dk} \left( k b-\frac{E\tau}{\hbar} +\theta(k) \right)=0 .
\label{spm_condition}
\end{equation}
This results in the following expression of the tunneling time,
\begin{equation}
\tau= \hbar \frac{d \theta(E)}{dE} +\frac{b}{(\frac{\hbar k}{m})} .
\label{phase_delay_time}
\end{equation}
For a square barrier potential $V(x)=V$ of width $b$, Eq. \ref{phase_delay_time} results in the following expression,
\begin{equation}
\tau= \hbar \frac{d}{dE} \tan^{-1} \left( \frac{k^{2}-q^{2}}{2kq} \tanh{q b}\right) .
\label{t_barrier}
\end{equation}
Here $q= \sqrt{2m(V-E)}/\hbar$. The following things are apparent from Eq. \ref{t_barrier}. 
\beq
\lim_{b \to 0} =0, \ \  \lim_{b \to \infty} =\frac{2m}{\hbar qk} .
\label{t_barrier_limit}
\eeq
The tunneling time is expected to vanish for $b \rightarrow 0$. However, the result for  $b \rightarrow \infty$ is highly unexpected as the tunneling time saturates to a finite value and is also independent of $b$. This shows that for thick barriers , the tunneling time is independent of the thickness of the barriers. This is the famous Hartman effect. We will use the system of unit  $2m=1$, $\hbar=1$, $c=1$ throughout the article. In this unit the tunneling time from the square barrier is,  
\begin{equation}
\lim_{b\rightarrow \infty} \tau =\frac{1}{qk}. 
\label{tt_qm}
\end{equation}
\section{Hartman effect from PT-symmetric barrier }
\label{hf_pt_discussions}
In this section we show that controversial Hartman effect exists for barriers arranged in $PT$-symmetric configurations. We first study the simplest $PT$-symmetric system made by the complex potential $u+iv$ and $u-iv$ each of thickness $b$ and arrange adjacently without  intervening gap. This is shown in Fig \ref{hf_simple_pt}.  We call this as `unit' PT-symmetric barrier system. Next we investigate the Hartman effect when this `unit' system repeats periodically to make a layered $PT$- symmetric barrier of finite repetition N as shown in Fig \ref{layered_pt_fig}. We also present our detailed calculations to show the $N \rightarrow \infty$ limit over a finite length $L$ of our layered $PT$-symmetric system gives the same tunneling time expression as of real rectangular barrier of height $u$ and length $L$. Therefore the Hartman effect from real barrier can be due to the Hartman effect of our layered $PT$-symmetric system in the special limiting case. For the purpose of clarity we discuss all the above three cases separately.      

\subsection{Unit PT-symmetric barrier }
\label{hf_simple_pt}
\begin{figure}
\begin{center}
\includegraphics[scale=0.5]{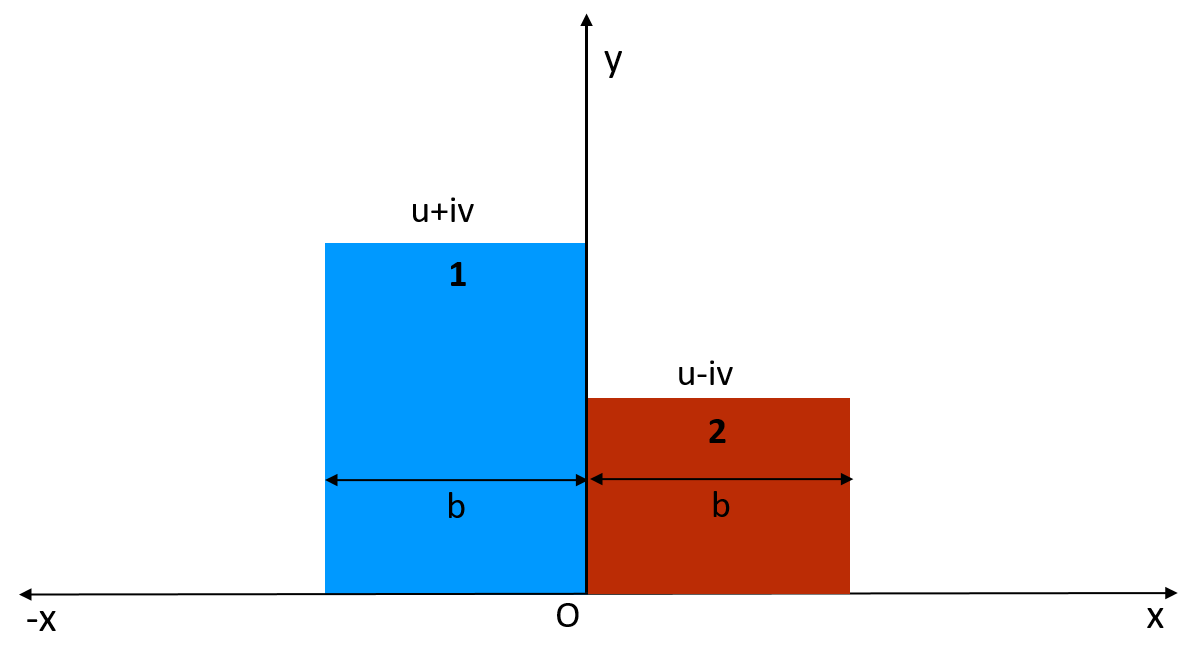}  
\caption{\it A $PT$-symmetric `unit cell' consisting a pair of complex conjugate barrier. $y$-axis represents the complex height of the potential.}  
\label{pt_fig}
\end{center}
\end{figure}  
In this section we calculate the tunneling time and investigate the Hartman effect from the following simple $PT$-symmetric system (shown in Fig- \ref{pt_fig})
\begin{eqnarray}
V(x)&=& u+iv  \ \ \ \ \ \mbox{for} \ -b<  x < 0  \nonumber  \\
V(x)&=& u-iv  \ \ \ \ \ \mbox{for} \  0 > x >b   \nonumber  \\
V(x)&=& 0     \ \ \ \ \ \mbox {for} \ |x| \geq b.   
\label{pt_potential}
\end{eqnarray} 
In the above $\{ u,v\} \in R^{+}$. It will be shown that the $PT$-symmetric potential given by  Eq. \ref{pt_potential}  display Hartman effect. The transmission coefficient ($t$)  for this potential can be easily calculated and given by
\beq
t=\frac{e^{i (\theta -2kb)}}{\sqrt{\xi^{2}+\chi^{2}}} , \ \ \ \theta= \tan^{-1} \left ( \frac{\chi}{\xi}\right ), 
\label{t_pt_system}
\eeq
and,
\beq
\xi=\frac{1}{2}(\cos{2\beta} +\cosh{2 \alpha}) + \cos{2\phi}(\cosh^{2}{\alpha} \sin^{2}{\beta} + \cos^{2}{\beta} \sinh^{2}{\alpha}) ,
\label{xi_exp}
\eeq
\beq
\chi=\frac{1}{2}(U_{+} \sin{\phi} \sin{2\beta} + U_{-} \cos{\phi} \sinh{2\alpha}) .
\label{chi_exp}
\eeq
In Eqs. \ref{xi_exp} and \ref{chi_exp}, the quantities $\alpha, \beta, \phi$ and $U_{\pm}$ are given by,
\beq
\alpha = b \rho \cos{\phi},\ \ \beta = b \rho \sin{\phi}, \ \ U_{\pm}= \frac{k}{\rho} \pm \frac{\rho}{k},  
\label{alpha_beta_u}
\eeq
and
\beq
\phi= \frac{1}{2} \tan^{-1} {\left ( \frac{v}{u- k^{2}}\right )}, \ \ \rho= \left[ (u-k^{2})^{2}+v^{2} \right]^{\frac{1}{4}} .
\label{rho_phi}
\eeq
Derivation of Eq. \ref{t_pt_system} is provided in Appendix- A (at the end) by transfer matrix approach. Using SPM, the tunneling time ($\tau$) is given by 
\beq
\tau=\frac{1}{2k} \left ( \frac{\xi \chi' -\chi \xi'}{\xi^{2}+ \chi^{2}} \right ). 
\label{t_unit_pt}
\eeq
We will use `$\prime$' (prime) to denote the derivatives with respect to wave vector $k$. The expressions for $\xi '$ and $\chi'$ are provided below
\beq
\xi'=2 \alpha'\cos^{2}{\phi} \sinh{2 \alpha} -2 \beta' \sin^{2}{\phi} \sin{2 \beta} + \phi' \sin{2 \phi} (\cos{2 \beta}-\cosh{2 \alpha}) .
\label{xi_prime} 
\eeq
\begin{multline}
\chi'= \frac{1}{2}\sin{\phi}( U_{+}' \sin{2 \beta} + 2\beta' U_{+} \cos{2 \beta} - \phi' U_{-} \sinh{2 \alpha}) +\\
\frac{1}{2} \cos{\phi}( U_{-}' \sinh{2 \alpha} +2 \alpha' U_{-} \cosh{2\alpha} +\phi' U_{+} \sin{2 \beta}) .
\label{chi_prime}
\end{multline}
The width dependency in tunneling time enters through $\alpha$ , $\alpha '$, $\beta$ and $\beta'$. Therefore for the existence of Hartman effect $\tau $ must be independent of these four quantities in the limit $b \rightarrow \infty$. It can be shown that
\beq
\lim_{b \rightarrow \infty } \xi =\frac{e^{2 \alpha}}{2} \cos^{2}{\phi} ,
\label{xi_limit}
\eeq 
\beq
\lim_{b \rightarrow \infty } \chi =\frac{U_{-}}{4}e^{2 \alpha} \cos{\phi} ,
\label{chi_limit}
\eeq 
\beq
\lim_{b \rightarrow \infty } \xi' = e^{2 \alpha} (\alpha' \cos^{2}{\phi} -\frac{\phi'}{2}  \sin{2 \phi} ) ,
\label{xi_prime_limit}
\eeq 
\beq
\lim_{b \rightarrow \infty } \chi' = \frac{e^{2 \alpha}}{4} \left ( \cos{\phi} (U_{-}' +2 \alpha' U_{-}) - U_{-} \phi' \sin{\phi} \right ) .
\label{chi_prime_limit}
\eeq 
Using the results of \ref{xi_limit}-\ref{chi_prime_limit} in Eq. \ref{t_unit_pt}, we find 
\beq
\lim_{b \rightarrow \infty} \tau = \tau_{\infty} =\frac{U_{-}' \cos{\phi}  + \phi' U_{-} \sin{\phi} }{k(4 \cos^{2}{\phi} + U_{-}^{2})} .
\label{hartman_unit_pt} 
\eeq
In the expression of $\tau_{\infty}$, $b$ dependent terms doesn't appear . This proves that the $PT$-symmetric system given by Eq. \ref{pt_potential}  shows Hartman effect.

\subsection{Layered PT-symmetric barrier}
\label{layered_section}
\begin{figure}
\begin{center}
\includegraphics[scale=0.65]{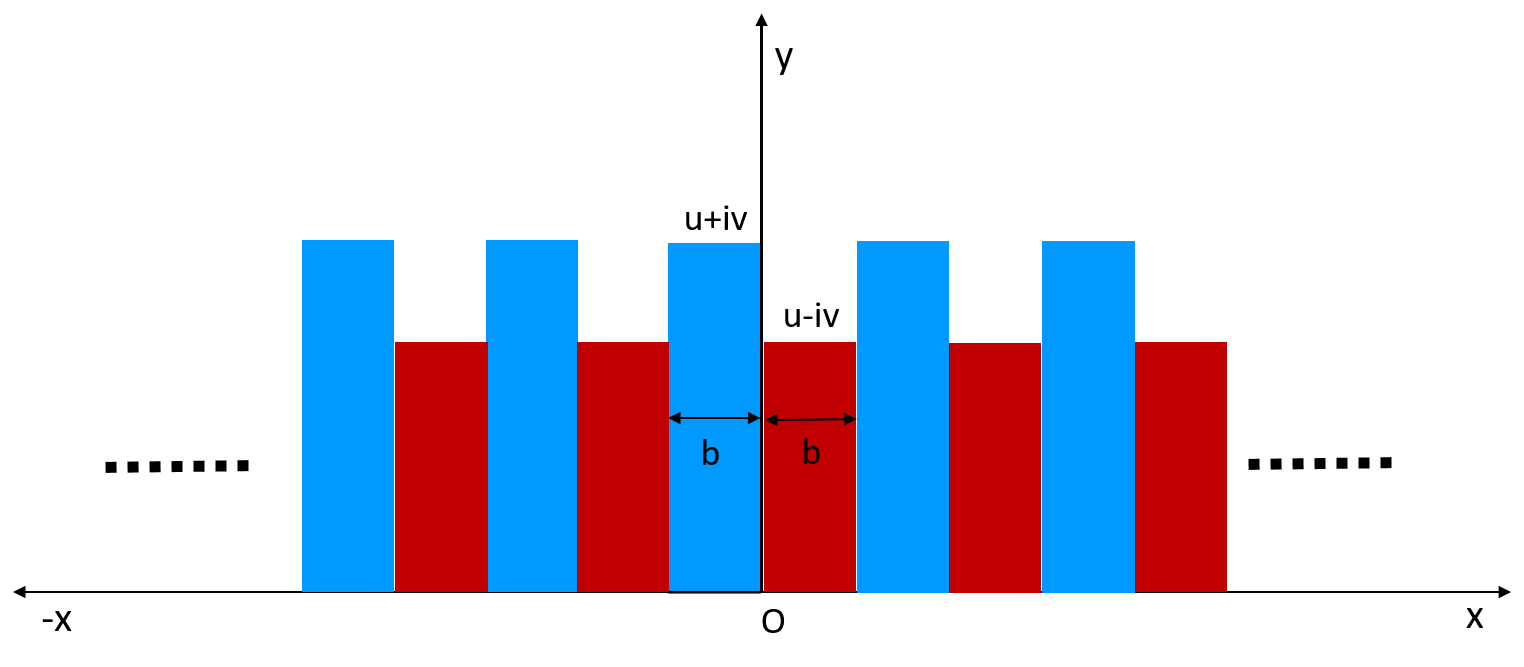}  
\caption{\it A locally periodic $PT$-symmetric system obtained by periodic repetitions of the `unit' $PT$-symmetric system shown in Fig \ref{pt_fig}. $y$-axis is the complex height of the potential.}  
\label{layered_pt_fig}
\end{center}
\end{figure} 
Next we calculate the tunneling time from a layered (locally periodic) $PT$-symmetric system obtained by periodic repetitions of the `unit cell' $PT$-symmetric system of Eq. \ref{pt_potential}. The layered $PT$-symmetric system is shown in Fig \ref{layered_pt_fig}. The net spatial extent of the layered $PT$-symmetric system is $L=2Nb$ where $N$ is the number of repetitions. It is easy to show that the transmission coefficient from such a system is (see Appendix-B for derivation)
\beq
t=\frac{e^{-ikL}}{H(k)} ,
\label{t_simple}
\eeq
where $H(k)$ is given as
\beq
H(k)=(\xi-i\chi ) U_{N-1}(\xi)-U_{N-2}(\xi) .
\eeq
The phase of the transmission coefficient is given by,
\beq
\Theta = \tan^{-1} ( g \chi   ) -kL ,
\label{phase_n}
\eeq
where,
\beq
g=\frac{U_{N-1} (\xi)}{T_{N} (\xi)} .
\label{g_eq}
\eeq
From the knowledge of the phase of the transmission coefficient, the tunneling time is calculated to be 
\beq
\tau_{N}= \frac{1}{2k(1+ g^{2}\chi^{2})} \left [g \chi' +  \chi \left ( \frac{N \xi'}{\xi^{2}-1} - N \xi' g^{2}- \frac{g \xi \xi'}{\xi^{2}-1}  \right ) \right ] .
\label{time_n}
\eeq
To show Hartman effect, we first evaluate the following limits, 
\beq
\lim _{b \rightarrow \infty} g \sim \frac{1}{\xi (b \rightarrow \infty)} ,
\label{g_infinite}
\eeq  
 \beq
\lim _{b  \rightarrow \infty} \frac{\chi}{\xi} = \frac{U_{-}}{2} \sec{\phi}=\eta .
\label{chi_xi_ratio_infinite}
\eeq    
Making use of the fact, $\lim_{b \rightarrow \infty} g \sim \frac{1}{\xi}$ and $\lim_{b \rightarrow  \infty} \xi >> 1$, we can write
\beq
\lim _{b \rightarrow \infty} \tau_{N}= \frac{1}{2k}  \left ( \frac{1}{1+ \eta^{2}} \right ) \left [ \lim_{b \rightarrow \infty} \left ( \frac{\chi'}{\xi}  -\frac{\chi \xi'}{\xi^{2}} \right ) \right ] .
\label{time_1}
\eeq
Using previously derived limits, the term in square parenthesis is given by
\beq
\lim_{b \rightarrow \infty} \left ( \frac{\chi'}{\xi}  -\frac{\chi \xi'}{\xi^{2}} \right ) = \frac{2}{\cos{\phi} } (U_{-}' +\phi U_{-} \tan{\phi}). 
\eeq
Therefore the limiting case of the tunneling time is turned out to be ,
\beq
\lim _{b \rightarrow \infty} \tau_{N}= \lim _{b \rightarrow \infty} \tau = \tau_{\infty}.  
\label{time_2}
\eeq
Eq. \ref{time_2} proves the Hartman effect from the layered PT-symmetric  system represented  by Fig \ref{layered_pt_fig} .  
\subsection {Real barrier}
\label{real_barrier_section}
In this sub-section we show that the famously known Hartman effect from a real barrier is the special limiting case of $PT$-symmetric system considered in the sub-section \ref{layered_section} above.  We first show that the transmission phase of a rectangular barrier of height $u$ and width $L$ is the limiting case of $N \rightarrow \infty$ of our layered $PT$-symmetric system such that $b=\frac{L}{2N}$ where $L$ is fixed. To prove this, we first Taylor expand the quantity `$g \chi$' in power of `$b$' such that,
\beq
g \chi =A_{0} + \sum_{j=1}^{\infty} A_{j} b^{j} . 
\eeq  
It is found that $A_{0}=0$ and the coefficients of even power of $b$ are also zero. Therefore
\beq
g \chi=  A_{1}  + A_{3} b^{3} + A_{5} b^{5} + A_{7} b^{7} + A_{9} b^{9} + ....
\label{g_taylor}
\eeq  
The coefficients of various powers of $b$ are given by,
\beq
A_{1}= N \rho (U_{-} \cos^{2}{\phi} + U_{+} \sin^{2}{\phi}) ,
\label{a1}
\eeq
\beq
A_{3}= - \frac{1}{6} N \rho^{3} \left [ 8 N^{2} (U_{+} + U_{-}) \cos{2 \phi} -(U_{+}-U_{-}) \{ 4 N^{2} -1 + (4 N^{2} +1) \cos{4 \phi} \} \right ] ,
\label{a3}
\eeq
\begin{multline}
A_{5}= \frac{N \rho ^{5}}{360} [ 2 (U_{+} + U_{-}) (96 N^{4} -5 N^{2} -1) - (U_{+} - U_{-}) \cos{2 \phi} (288 N^{4} -25 N^{2} -8) + \\  2 (U_{+} + U_{-}) \cos{4 \phi} (96 N^{4} + 5 N^{2} +1) - (U_{+} - U_{-}) \cos{6 \phi} (96 N^{4} + 25 N^{2} +8)  ] ,
\label{a5}
\end{multline}
\begin{multline}
A_{7}= \frac{N \rho ^{7}}{15120} [  (U_{+} - U_{-}) (9792 N^{6} -1008 N^{4} -161 N^{2} -34) -\\ 4 (U_{+} + U_{-}) \cos{2 \phi} (4896 N^{6} -168 N^{4} -35 N^{2} -10) + \\ 4 (U_{+} - U_{-}) \cos{4 \phi} (3264 N^{6} -35 N^{2} -16) - \\ 4 (U_{+} + U_{-}) \cos{6 \phi} (1632 N^{6} + 168 N^{4} + 35 N^{2} +10) + \\ (U_{+} - U_{-}) \cos{8 \phi} (3264 N^{6} + 1008 N^{4} + 301 N^{2} +98)   ] ,
\label{a7}
\end{multline}
\begin{multline}
A_{9}= \frac{N \rho ^{9}}{453600} [ 4 (U_{+} + U_{-}) (119040 N^{8} - 6120 N^{6} -1029 N^{4} -215 N^{2} -61) - \\ 2 (U_{+} - U_{-}) \cos{2 \phi} (396800 N^{8} -28560 N^{6} -5502 N^{4} - 1045 N^{2} -268) + \\ 40 (U_{+} + U_{-}) \cos{4 \phi} ( 15872 N^{8} -42 N^{4} -15 N^{2} -5) - \\ (U_{+} - U_{-}) \cos{6 \phi} (396800 N^{8} + 28560 N^{6} + 1722 N^{4} - 655 N^{2} -352) + \\ 4 (U_{+} + U_{-}) \cos{8 \phi} (39680 N^{8} + 6120 N^{6} + 1449 N^{4} + 365 N^{2} + 111) - \\ (U_{+} - U_{-}) \cos{10 \phi} (79360 N^{8} + 28560 N^{6} + 9282 N^{4} + 2745 N^{2} + 888)   ].
\label{a9}
\end{multline}
Similarly other coefficients $A_{11}, A_{13} , A_{15}$ etc. can also be calculated. Next we take the limiting case  $N \rightarrow \infty$ of these coefficients to show, 
\beq
\lim_{N \to \infty} A_{1} b= \left ( \frac{k^{2}-q^{2}}{2kq} \right ) (q L) ,
\label{a1b}
\eeq
\beq
\lim_{N \to \infty} A_{3} b^{3}= - \frac{1}{3} \left ( \frac{k^{2}-q^{2}}{2kq} \right )  (q L)^{3} ,
\label{a3b}
\eeq
\beq
\lim_{N \to \infty} A_{5} b^{5}= \frac{2}{15} \left ( \frac{k^{2}-q^{2}}{2kq} \right )  ( q L)^{5} ,
\label{a5b}
\eeq
\beq
\lim_{N \to \infty} A_{7} b^{7}= - \frac{17}{315} \left ( \frac{k^{2}-q^{2}}{2kq} \right )  ( q L)^{7} ,
\label{a7b}
\eeq
\beq
\lim_{N \to \infty} A_{9} b^{9}=  \frac{31}{2835} \left ( \frac{k^{2}-q^{2}}{2kq} \right )  ( q L)^{9}.
\label{a9b}
\eeq
Where on deriving Eqs. \ref{a1b}-\ref{a9b}, we have identified $L=2 Nb$ and expanded $\cos{4\phi}, \cos{6\phi}, \cos{8\phi}$ and $\cos{10\phi}$ in power of $\cos{2 \phi}= \frac{q^{2}}{\rho^{2}}$. The detail calculations of deriving Eqs. \ref{a1b}-\ref{a9b} are shown in Appendix-C . Thus,  
\beq
\lim_{N \to \infty , b \to 0} g \chi =  \left ( \frac{k^{2}-q^{2}}{2kq} \right ) \left [ q L - \frac{1}{3} (q L)^{3} + \frac{2}{15}(q L)^{5} - \frac{17}{315}(q L)^{7} +  \frac{31}{2835}   (q L)^{9} - ....  \right ].
\eeq
And we identify  
\beq
\lim_{N \to \infty , b \to 0} g \chi =  \left ( \frac{k^{2}-q^{2}}{2kq} \right ) \tanh{qL} , \ \ L=2Nb
\label{gchi_limiting_case}
\eeq
Using Eq. \ref{gchi_limiting_case} in Eq. \ref{phase_n} we find
\beq
\lim_{N \to \infty , b \to 0}  \Theta =  \tan^{-1} \left ( \frac{k^{2}-q^{2}}{2kq} \tanh{qL} \right  ) -k L. 
\eeq
Thus the tunneling time, 
\beq
\lim_{N \to \infty , b \to 0}  \tau_{N} = \frac{d}{dE} \left [ \tan^{-1} \left ( \frac{k^{2}-q^{2}}{2kq} \tanh{qL} \right  ) \right ].
\label{taun_limit} 
\eeq
Eq. \ref{taun_limit} yield the same value of tunneling time as Eq. \ref{tt_qm} . This result shows that Hartman effect in real barrier occur due to the limiting case  $ N \rightarrow \infty$ of our layered $PT$-symmetric system such that each layered structure becomes infinitely thin. It can be shown that a real barrier of height $u$ and width $L$ is the limiting case $N \rightarrow \infty$ of our layered $PT$-symmetric system such that   $b =\frac{L}{2 N}$ (we have left the complete exercise, however one can show that it exactly gives the expression of reflection and transmission coefficient of rectangular barrier. We have checked this numerically also) . Here $L$ is the net spatial extent of the layered $PT$-symmetric system.  
\section {Non PT-symmetric barrier system: No Hartman effect}
\label{non_pt_section}
\begin{figure}
\begin{center}
\includegraphics[scale=0.5]{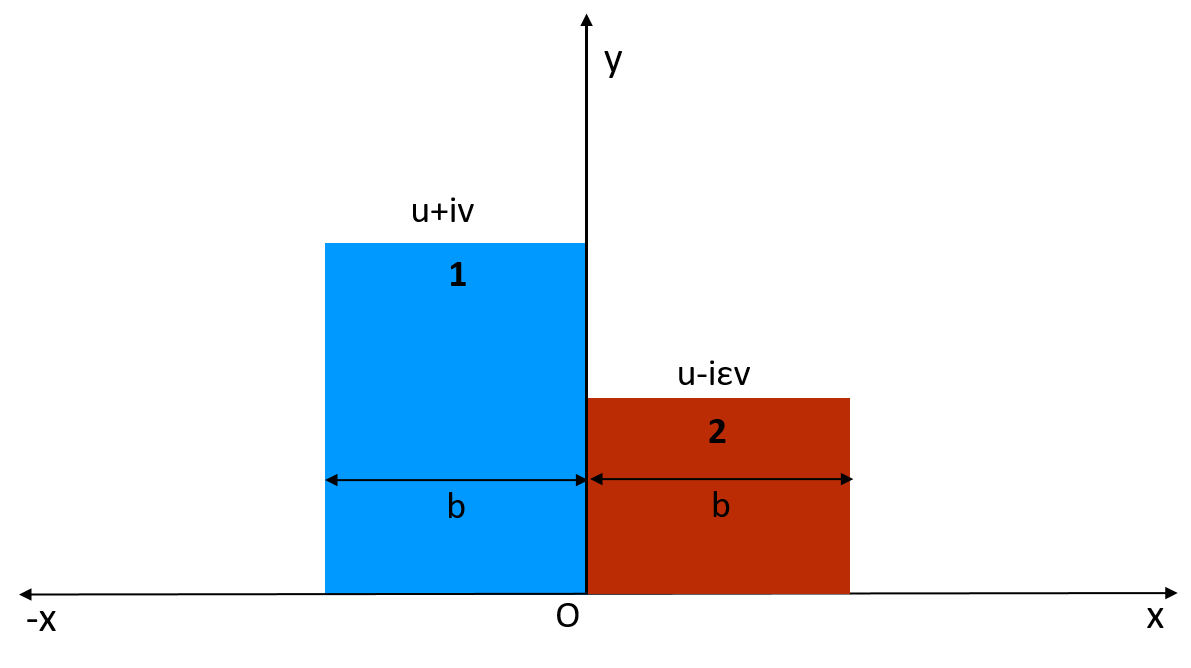}  
\caption{\it A non-Hermitian `unit cell' consisting a pair of complex conjugate barrier. $y$-axis represents the complex height of the potential. Note that the system becomes $PT$-symmetry when $\varepsilon =1$ and in this case identical to the `unit cell' shown in the Fig \ref{pt_fig}. }  
\label{pt_fig_eps}
\end{center}
\end{figure}

In this section we calculate the tunneling time by SPM method from the following  non-Hermitian system 
\begin{eqnarray}
V(x)&=& u+iv,   \ \ \ \ \ \mbox{for} \ -b<  x < 0.  \nonumber  \\
V(x)&=& u-i \varepsilon v ,  \ \ \ \ \ \mbox{for} \  0 > x >b .   \nonumber  \\
V(x)&=& 0 ,     \ \ \ \ \ \mbox {for} \ |x| \geq b.   
\label{non_pt_potential}
\end{eqnarray}
Where we have $\varepsilon \in R $. The system is shown graphically in Fig \ref{pt_fig_eps}. When $\varepsilon  =1$ , the non-Hermitian system represented by Eq. \ref{non_pt_potential} is $ PT $-symmetric and is identical to the system represented by Eq. \ref{pt_potential}. The transmission coefficient from this system can be calculated and is given by
\beq
t=\frac{e^{-2ikb}}{Q}.
\label{t_eps}
\eeq    
Where, 
\beq
Q=P^{-}_{1} P^{-}_{2} - S_{1}S_{2},
\eeq
and the various symbols are given by
\beq
P^{\pm}_{1,2}= 2 \cos{k_{1,2}b} \pm i \left ( \mu_{1,2} + \frac{1}{\mu_{1,2}} \right ) \sin{k_{1,2}b} ,
\eeq
\beq
S_{1,2}= i \left ( \mu_{1,2} - \frac{1}{\mu_{1,2}} \right ) \sin{k_{1,2}b}. 
\eeq
Here,
\beq
\mu_{1,2} = \frac{k_{1,2}}{k} ,
\eeq
and,
\beq
k_{1,2} =\sqrt{E- V_{1,2}},  \  V_{1}= u+iv , \ V_{2}= u-i \varepsilon v . 
\eeq
We express $k_{1} = \rho_{1} e^{i \phi_{1}}$, $k_{2} = \rho_{2} e^{-i \phi_{2}}$ and define the following quantities, 
\beq
H_{\pm} = \frac{\rho_{1}\rho_{2}}{k^{2}} \pm \frac{k^{2}}{ \rho_{1}\rho_{2}},  \ \ \ G_{\pm} = \frac{\rho_{1}}{\rho_{2}} \pm  \frac{\rho_{2}}{\rho_{1}} .
\eeq 
\beq
J_{1,2}^{\pm} = \frac{\rho_{1,2}}{k^{2}} \pm \frac{k^{2}}{\rho_{1,2}} .
\eeq
\beq
\rho_{1} = [(u-k^{2})^{2}+ v^{2}]^{\frac{1}{4}},  \ \ \  \rho_{2} = [(u-k^{2})^{2}+ \epsilon^{2}v^{2}]^{\frac{1}{4}}.
\eeq
\beq
\phi_{1} = \frac{1}{2} \tan^{-1} \left ( \frac{v}{u-k^{2}} \right ) , \ \ \ \phi_{2} = \frac{1}{2} \tan^{-1} \left ( \frac{ \epsilon v}{u-k^{2}} \right ) .
\eeq
Through the use of the above quantities, we separate $Q$ in real and imaginary parts (to obtain the phase of transmission coefficient given by Eq. \ref{t_eps}) 
\beq
Q= (A_{1} - A_{2}) + i (B_{1} - B_{2}).
\eeq
Where,
\begin{multline}
A_{1} =4 z_{1} + 2 (x_{2} J_{2}^{+} \cos{\phi_{2}} -x_{1} J_{2}^{-} \sin{\phi_{2}} ) + 2 (w_{1} J_{1}^{-} \sin{\phi_{1}} +  w_{2} J_{1}^{+} \cos{\phi_{1}} ) - \\ y_{1}(H_{+} \cos{(\phi_{1} -\phi_{2} )} + G_{+} \cos{(\phi_{1} +\phi_{2} )} ) + y_{2}(H_{-} \sin{(\phi_{1} -\phi_{2} )} + G_{-} \sin{(\phi_{1} +\phi_{2} )} ) 
\end{multline}  
\begin{multline}
B_{1} =4 z_{2} - 2 (x_{1} J_{2}^{+} \cos{\phi_{2}} +x_{2} J_{2}^{-} \sin{\phi_{2}} ) - 2 (w_{1} J_{1}^{+} \cos{\phi_{1}} -  w_{2} J_{1}^{-} \sin{\phi_{1}} ) - \\ y_{1}(H_{-} \sin{(\phi_{1} -\phi_{2} )} + G_{-} \sin{(\phi_{1} +\phi_{2} )} ) - y_{2}(H_{+} \cos{(\phi_{1} -\phi_{2} )} + G_{+} \cos{(\phi_{1} +\phi_{2} )} ) 
\end{multline} 
\begin{equation}
A_{2} = y_{2}(H_{-} \sin{(\phi_{1} -\phi_{2} )} - G_{-} \sin{(\phi_{1} +\phi_{2} )} ) - y_{1}(H_{+} \cos{(\phi_{1} -\phi_{2} )} - G_{+} \cos{(\phi_{1} +\phi_{2} )} ) 
\end{equation} 
\begin{equation}
B_{2} = -y_{2}(H_{+} \cos{(\phi_{1} -\phi_{2} )} - G_{+} \cos{(\phi_{1} +\phi_{2} )} ) - y_{1}(H_{-} \sin{(\phi_{1} -\phi_{2} )} - G_{-} \sin{(\phi_{1} +\phi_{2} )} ) 
\end{equation} 
In the above, the quantities  $w_{1,2}, x_{1,2}, y_{1,2}, z_{1,2}$ are due to,
\beq
\sin{k_{1}b} \cos{k_{2}b} = w_{1}+ i w_{2} , \ \      \cos{k_{1}b} \sin{k_{2}b} = x_{1}+ i x_{2} , 
\eeq
\beq
\sin{k_{1}b} \sin{k_{2}b} = y_{1}+ i y_{2} , \ \      \cos{k_{1}b} \cos{k_{2}b} = z_{1}+ i z_{2} . 
\eeq
The expressions of $w_{1,2}, x_{1,2}, y_{1,2}, z_{1,2}$ are given below,
\begin{eqnarray}
w_{1} & =& \cos{\alpha_{22}} \cosh{\beta_{11}} \cosh{\beta_{22}}\sin{\alpha_{11}}  - \cos{\alpha_{11}} \sin{\alpha_{22}} \sinh{\beta_{11}} \sinh{\beta_{22}} , \\
w_{2} & =&  \cos{\alpha_{11}} \cos{\alpha_{22}}  \cosh{\beta_{22}} \sinh{\beta_{11}} + \cosh{\beta_{11}} \sin{\alpha_{11}} \sin{\alpha_{22}}  \sinh{\beta_{22}} , \\
x_{1} & =& \cos{\alpha_{11}} \cosh{\beta_{11}} \cosh{\beta_{22}}\sin{\alpha_{22}}  - \cos{\alpha_{22}} \sin{\alpha_{11}} \sinh{\beta_{11}} \sinh{\beta_{22}} ,\\
x_{2} & =& - \cosh{\beta_{22}} \sin{\alpha_{11}} \sin{\alpha_{22}} \sinh{\beta_{11}} - \cos{\alpha_{11}} \cos{\alpha_{22}} \cosh{\beta_{11}}  \sinh{\beta_{22}} , \\
y_{1} & =& \cosh{\beta_{11}} \cosh{\beta_{22}}\sin{\alpha_{11}} \sin{\alpha_{22}} + \cos{\alpha_{11}} \cos{\alpha_{22}} \sinh{\beta_{11}} \sinh{\beta_{22}} ,\\
y_{2} & =& \cos{\alpha_{11}} \cosh{\beta_{22}} \sin{\alpha_{22}} \sinh{\beta_{11}} - \cos{\alpha_{22}} \cosh{\beta_{11}} \sin{\alpha_{11}} \sinh{\beta_{22}} ,   \\
z_{1} &= &\cos{\alpha_{11}} \cos{\alpha_{22}} \cosh{\beta_{11}} \cosh{\beta_{22}} + \sin{\alpha_{11}} \sin{\alpha_{22}} \sinh{\beta_{11}} \sinh{\beta_{22}} ,\\
z_{2} &=& -\cos{\alpha_{22}} \cosh{\beta_{22}} \sin{\alpha_{11}} \sinh{\beta_{11}} + \cos{\alpha_{11}} \cosh{\beta_{11}} \sin{\alpha_{22}} \sinh{\beta_{22}} . 
 \end{eqnarray}
In the above
\beq
\alpha_{ij}=b \rho_{i} \cos{\phi_{j}}, \ \ \ \beta_{ij}=b \rho_{i} \sin{\phi_{j}} .
\eeq
Now, the phase of transmission coefficient can be found as,
\begin{equation}
\theta = \Phi_{\varepsilon } -2kb. 
\end{equation}
Where we have ,
\beq
\Phi_{\varepsilon } = \tan^{-1}{\left (\frac{B_{2} - B_{1}}{ A_{1} -A_{2}} \right )}  .
\eeq
Thus the tunneling time is ,
\begin{equation}
\tau_{\varepsilon} = \frac{d}{dE} (\Phi_{\varepsilon } -2kb) + \frac{2b}{2k} .
\end{equation}
The last term of R.H.S. in the above equation is due to the free propagation time of traversing the length $2b$. The net tunneling time can be written as 
\begin{equation}
\tau_{\varepsilon} = \frac{d \Phi_{\varepsilon}}{dE} =\frac{1}{2k} \frac{d \Phi_{\varepsilon}}{dk}  .
\end{equation}
In order to analyze the effect of $PT-$ symmetry over Hartman effect, we Taylor expand $\tau_{\varepsilon }$ near $\varepsilon \sim 1$ to first order as follows ,
\begin{equation}
\tau_{\varepsilon} =    \tau_{\varepsilon} (\varepsilon =1) + \left (\frac{d \tau_{\varepsilon}}{d \varepsilon} \right )_{\varepsilon =1} (\varepsilon -1) .
\label{tau_eps_taylor_expand}
\end{equation}
We take the limit $\lim_{ b \rightarrow \infty} $ of Eq. \ref{tau_eps_taylor_expand} to study Hartman effect near the symmetry breaking threshold $\varepsilon =1$. Taking the limit of Eq. \ref{tau_eps_taylor_expand},
\begin{equation}
\lim_{b \to \infty} \tau_{\varepsilon} = \lim_{b \to \infty}   \tau_{\varepsilon} ( \varepsilon =1) + \lim_{b \to \infty} \left (\frac{d \tau_{\varepsilon}}{d \varepsilon} \right )_{\varepsilon =1} (\varepsilon -1) .
\end{equation}
The first term of right hand side is $\tau_{\infty}$ and is independent of $b$ . Thus ,    
\begin{equation}
\lim_{b \to \infty} \tau_{\varepsilon} = \tau_{\infty} + \lim_{b \to \infty} \left (\frac{d \tau_{ \varepsilon}}{d \varepsilon} \right )_{\varepsilon =1} (\varepsilon -1) .
\end{equation}
Therefore to find whether Hartman effect exist or not when $PT-$ symmetry is broken, we investigate the second term of R.H.S about its dependency on the thickness $b$ in the limit $b \rightarrow \infty$. We first evaluate the following derivative in the limit $b \rightarrow \infty$, 
\begin{equation}
\lim_{b \to \infty} \left (\frac{d \tau_{\varepsilon}}{ d\varepsilon} \right )=  \frac{1}{2k}  \lim_{b \to \infty} \left ( \frac{d}{d \varepsilon } (\frac{d \Phi_{\varepsilon }}{dk}) \right )
\end{equation}
As $\varepsilon $ , $k$ and $b$ are independent quantities, we can take the limit inside the differential sign. Thus we write,
\begin{equation}
\lim_{b \to \infty} \left (\frac{d \tau_{\varepsilon}}{ d\varepsilon} \right )=  \frac{1}{2k}   \left [ \frac{d}{d \varepsilon } \left (\frac{d }{dk} (\lim_{b \to \infty} \Phi_{\varepsilon } ) \right ) \right ] .
\label{tau_ep_lim}
\end{equation}
In the next we evaluate $\lim_{b \to \infty} \Phi_{\varepsilon }$. For this we evaluate the limiting values of the following quantities,

\begin{eqnarray}
\lim_{b \to \infty} z_{1}  & = &  \frac{1}{4} e^{\beta_{11} +\beta_{22}} \cos{(\alpha_{11} -\alpha_{22})} ,\\
\lim_{b \to \infty} z_{2}  & = &  \frac{1}{4} e^{\beta_{11} +\beta_{22}} \sin{(\alpha_{22} -\alpha_{11})} ,\\
\lim_{b \to \infty} y_{1}  & = &  \frac{1}{4} e^{\beta_{11} +\beta_{22}} \cos{(\alpha_{11} -\alpha_{22})} ,\\
\lim_{b \to \infty} y_{2}  & = &  \frac{1}{4} e^{\beta_{11} +\beta_{22}} \sin{(\alpha_{22} -\alpha_{11})} ,\\
\lim_{b \to \infty} x_{1}  & = &  \frac{1}{4} e^{\beta_{11} +\beta_{22}} \sin{(\alpha_{22} -\alpha_{11})} ,\\
\lim_{b \to \infty} x_{2}  & = &  -\frac{1}{4} e^{\beta_{11} +\beta_{22}} \cos{(\alpha_{11} -\alpha_{22})} ,\\
\lim_{b \to \infty} w_{1}  & = &  \frac{1}{4} e^{\beta_{11} +\beta_{22}} \sin{(\alpha_{11} -\alpha_{22})} ,\\
\lim_{b \to \infty} w_{2}  & = &  \frac{1}{4} e^{\beta_{11} +\beta_{22}} \cos{(\alpha_{11} -\alpha_{22})} .
\end{eqnarray}
It is observe that, 
\begin{eqnarray}
\lim_{b \to \infty} z_{1}  & = &  \lim_{b \to \infty} y_{1} , \\
\lim_{b \to \infty} z_{2}  & = &  \lim_{b \to \infty} y_{2} , \\
\lim_{b \to \infty} x_{1}  & = &  \lim_{b \to \infty} y_{2} , \\
\lim_{b \to \infty} x_{2}  & = &  -\lim_{b \to \infty} y_{1} , \\
\lim_{b \to \infty} w_{1}  & = &  -\lim_{b \to \infty} y_{2} , \\
\lim_{b \to \infty} w_{2}  & = &  \lim_{b \to \infty} y_{1} .
\end{eqnarray}
We define,
\begin{equation}
Y_{1} = \lim_{b \to \infty} y_{1} \ , Y_{2} = \lim_{b \to \infty} y_{2} ,
\end{equation}
so that,
\begin{equation}
Y_{2} =  -Y_{1} \tan{(\alpha_{11} -\alpha_{22})} .
\end{equation}
With these results we simplify the expression of $\Phi_{\varepsilon }$ in the limit $b \rightarrow \infty$ to obtain,
\begin{equation}
\lim_{b \to \infty} \Phi_{\varepsilon } = \tan^{-1} \left ( \frac{Q_{1} \tan{\zeta -Q_{2}}}{ Q_{1} + Q_{2} \tan{\zeta}} \right ).
\end{equation}
Where, $Q_{1}$ and $Q_{2}$ are given by
\beq
Q_{1} = 2 + J_{1}^{+} \cos{\phi_{1}} - J_{2}^{+} \cos{\phi_{2}} - G_{+} \cos{(\phi_{1}+\phi_{2})},
\eeq
\beq
Q_{2} =  J_{1}^{-} \sin{\phi_{1}} + J_{2}^{-} \sin{\phi_{2}} - G_{-} \sin{(\phi_{1}+\phi_{2})}.
\eeq
For future calculations in mind, we define the quantity $P$ as,
\beq
P= \frac{Q_{1} \tan{\zeta -Q_{2}}}{ Q_{1} + Q_{2} \tan{\zeta}}.
\label{p_exp}
\eeq
so that, 
\begin{equation}
\lim_{b \to \infty} \Phi_{\varepsilon } = \tan^{-1} P.
\label{phi_ep_lim}
\end{equation}
Using Eq. \ref{phi_ep_lim} in Eq. \ref{tau_ep_lim}, we find the following expression at $\varepsilon =1$, 
\begin{equation}
\left [ \lim_{b \to \infty} \left (\frac{d \tau_{\varepsilon}}{ d\varepsilon} \right ) \right ]_{\varepsilon =1}=    \frac{1}{2k}   \left [ \frac{1}{(1+ P^{2})}  \frac{d^{2}P}{d \varepsilon dk} - \frac{2P}{(1+ P^{2})^{2}} \frac{dP}{dk} \frac{dP}{d \varepsilon}\right ]_{\varepsilon =1} .
\label{tau_eps1}
\end{equation}
It is a massive calculation to evaluate the right hand side of Eq. \ref{tau_eps1}. The details of the  calculations are provided in Appendix-D. The term in the parenthesis of the right hand side is given by, 
\beq
\left [ \frac{1}{(1+ P^{2})}  \frac{d^{2}P}{d \varepsilon dk} - \frac{2P}{(1+ P^{2})^{2}} \frac{dP}{dk} \frac{dP}{d \varepsilon}\right ]_{\varepsilon =1} = K_{0} + K_{1} b.
\eeq 
The expressions for $K_{0}$ and $K_{1}$ are provided in the Appendix-D . Both  $K_{0}$ and $K_{1}$, are independent of `$b$'. Now, the net tunneling time in the vicinity of $\varepsilon \sim 1$ for large thickness `$2b$' is ,
\beq
\lim_{b \to \infty} \tau_{\varepsilon} =  \left [\tau_{\infty} +  \frac{K_{0}}{2k}(\varepsilon -1) \right ] +  \frac{K_{1}}{2k}(\varepsilon -1)b.
\label{tau_eps2}
\eeq 
It is clear from the above Eq. \ref{tau_eps2} that the tunneling time depends upon the thickness when $\varepsilon \neq 1$ i.e. Hartman effect is lost when the $PT$-symmetry is broken. It is easily seen from Eq. \ref{tau_eps2} that Hartman effect is restored when $\varepsilon = 1$ i.e when the system recovers the $PT$-symmetry.

\section{Conclusions and Discussions}
\label{results_discussions}
We have investigated the role of $PT$-symmetry for the occurrence of Hartman effect. We have considered a `unit cell'  $PT$-symmetric potential made by the two potentials of height $u+iv$ and $u-iv$ without an inter barrier separation and each of equal width `$b$'. We found that when $b \rightarrow \infty$, the tunneling time saturates and become independent of $b$. Thus Hartman effect exist in this $PT$-symmetric potential. Further it was found that layered $PT$-symmetric potential made by an arbitrary $N$ repetitions of this `unit cell' potential also shows Hartman effect. We have analytically investigated the case of infinite $N$ repetitions over finite spatial length $L$ and found that $N \rightarrow \infty$ limit results in the  same analytical expression of tunneling time as that of rectangular barrier of height `$u$' and width $L$. This result shows that the Hartman effect from a real barrier can be due to the Hartman effect from our layered $PT$-symmetric system. Also, the real rectangular barrier of height $u$ and $L$ is the limiting case $N \rightarrow \infty$ of our layered $PT$-symmetric system over fixed spatial length $L$.
     
To study the occurrence of Hartman effect at symmetry breaking threshold, we consider the tunneling time through a non-Hermitian potential made by two potentials of height  $u+iv$ and $u- i\varepsilon v$ each of equal width `$b$'. Expression of tunneling time is obtained analytically at the symmetry breaking threshold $\varepsilon \sim  1$ . It is found that when $ \varepsilon \neq 1 $, i.e. when $PT$-symmetry is broken, the Hartman effect is lost from the system. However, when   $ \varepsilon = 1 $ i.e. when $PT$-symmetry is respected, the Hartman effect is restored. This result along with the result of our layered $PT$-symmetric system and its limiting case $N \rightarrow \infty$ for fixed $L$ indicates that  $PT$-symmetry could be playing an important role for the occurrence of Hartman effect. 

{\it \bf{Acknowledgements}}: \\
MH acknowledges supports from SSPO for the encouragement of research activities. BPM acknowledges the support from MATRIX project (Grant No. MTR/2018/000611),
SERB, DST Govt. of India. .

\begin{center}
	{\large \bf Appendix - A : Derivation of transmission coefficient from unit PT-symmetric barrier }
\end{center}
The transfer matrices for the two barriers `$1$' and `$2$' as labeled in the Fig \ref{pt_fig} are given by 
\beq
 M_{1,2}(k)= \frac{1}{2 }\begin{pmatrix}   e^{-ikb} P_{+}^{1,2} & e^{-ikb(1+2j)} S^{1,2} \\ -e^{ikb(1+2j)} S^{1,2} & e^{ikb} P_{-}^{1,2}   \end{pmatrix} . 
\eeq 
In the above matrix, $j=0$ for barrier-$1$ and $j=1$ for barrier-$2$. Various symbols are given below,
\beq
P_{\pm}^{1,2}=2 \cos{k_{1,2} b} \pm i \left(\mu_{1,2} +\frac{1}{\mu_{1,2}} \right) \sin{k_{1,2} b} ,
\label{p_expression}
\eeq
\beq
S^{1,2}= i \left(\mu_{1,2} -\frac{1}{\mu_{1,2}} \right) \sin{k_{1,2} b} ,
\label{s_expression}
\eeq
\beq
\mu_{1,2}=\frac{k_{1,2}}{k}, \ \  k_{1,2}=\sqrt{k^{2}-V_{1,2}} .
\label{u12}
\eeq
For the potential represented by Eq. \ref{pt_potential} (or by Fig \ref{pt_fig}) $V_{1} = u+ iv$ and $V_{2} =u- iv$. From the composition properties of the transfer matrix, we can find the net transfer matrix $M$, of our $PT$-symmetric `unit cell'  as $M(k)=M_{2}(k). M_{1}(k)$. Therefore, 
\beq
 M(k)= \frac{1}{4 }\begin{pmatrix}   e^{-2ikb} (P_{+}^{1} P_{+}^{2}-S^{1}S^{2}) & e^{-2ikb} (P_{+}^{2} S^{1}+P_{-}^{1} S^{2}) \\ -e^{2ikb} (P_{-}^{2} S^{1}+P_{+}^{1} S^{2}) & e^{2ikb} (P_{-}^{1} P_{-}^{2}-S^{1}S^{2})   \end{pmatrix} .
\label{pt_transfer} 
\eeq  
Now the transmission coefficient (inverse of the $M_{22}$ element ) can be expressed as, 
\beq
t= \frac{e^{-2ikb}}{(P_{-}^{1} P_{-}^{2}-S^{1}S^{2})} .
\label{t_unitpt_appendix}
\eeq
We separate the denominator in real and imaginary parts. To do this  we first express $k_{1} =\sqrt{k^{2}- (u+iv)} = \rho e^{i \phi}$ and $k_{2} =\sqrt{k^{2}- (u-iv)} = \rho e^{-i \phi}$ where,
\beq
\rho =[ (u-k^{2})^{2}+ v^{2}]^{\frac{1}{4}} , \ \ \phi= \frac{1}{2} \tan^{-1} \left ( \frac{v}{u -k^{2}}\right ).
\eeq
Upon substituting $k_{1,2}$ expressions, the denominator of Eq. \ref{t_unitpt_appendix} is simplified to
\beq
P_{-}^{1} P_{-}^{2}-S^{1}S^{2} = \xi -i \chi = (\sqrt{\xi^{2} + \chi^{2}}) e^{-i \theta},
\label{d_simpl_appendix}
\eeq
where,
\beq
\theta = \tan^{-1}{\left ( \frac{\chi}{\xi}\right )}
\eeq
$\xi $ and $\chi $ are given by Eq. \ref{xi_exp} and \ref{chi_exp} respectively. Substitution of Eq. \ref{d_simpl_appendix} in Eq. \ref{t_unitpt_appendix} leads to
\beq
t= \frac{e^{i (\theta -2kb})}{\sqrt{\xi^{2} + \chi^{2}}} .
\eeq
This is Eq. \ref{t_pt_system}. 
\\
\begin{center}
	{\large \bf Appendix - B : Derivation of the transmission coefficient from finite layered PT-symmetric barrier }
\end{center}
If the transfer matrix $M$, 
\beq
 M(k)= \begin{pmatrix}   M_{11}(k) & M_{12}(k) \\ M_{21}(k) & M_{22}(k)   \end{pmatrix} , 
\eeq
of a `unit cell' potential is known such that, 
\beq
\begin{pmatrix}   A_{+}(k) \\ B_{+}(k)     \end{pmatrix}= M(k) \begin{pmatrix}   A_{-}(k) \\ B_{-}(k)    \end{pmatrix} .
\label{t_matrix}
\eeq
Where coefficients of the asymptotic solution of the scattering wave to the right of the potential are $A_{+}, B_{+}$ and to the left of the potential are $A_{-}, B_{-}$ . Then the transmission coefficient (for incidence from left) of a periodic system made by $n$ repetitions of the `unit cell' is given by
\beq
t_{n}= \frac{e^{-i k n s}}{[M_{22}(k) e^{-iks} U_{n-1}(\Omega ) - U_{n-2} ( \Omega )] }.
\label{tn_eq}
\eeq
Where,
\beq
\Omega= \frac{1}{2} (M_{11}e^{iks} +M_{22} e^{-iks}),
\label{omega_expression}
\eeq
with $s=w+g$. Here $w$ is width of the `unit cell' potential and $g$ is the gap between consecutive `unit cell' potentials . For our present problem (section \ref{layered_section}), $w=2b$ and $g=0$, thus $s=2b$. The procedure to derive Eq. \ref{tn_eq}  is outlined in \cite{griffith_periodic} . 
$M_{11}$ and $M_{22}$ elements of our `unit cell' potential are given in Eq. \ref{pt_transfer}. Using Eq. \ref{d_simpl_appendix},  $M_{22}$ element can be written as 
\beq
M_{22}=  (\xi -i\chi) e^{2ikb}.
\label{m22_simp} 
\eeq
Similarly,
\beq
M_{11}=  (\xi +i\chi) e^{-2ikb}.
\label{m11_simp} 
\eeq
To arrive at Eq. \ref{m11_simp}, we have separated term $P_{+}^{1} P_{+}^{2}-S^{1}S^{2}$ in real and imaginary parts. The expression for $\xi$  and $\chi$ are given by Eq. \ref{xi_exp} and Eq. \ref{chi_exp} respectively. From simplified expressions of $M_{11}$ and $M_{22}$ , we observe $M_{22}=M_{11}^{*}$. This shows that the argument, $\Omega$ of Chebyshev polynomial is real.  We substitute Eq. \ref{m22_simp} and Eq. \ref{m11_simp} in Eq. \ref{omega_expression} to obtain  $\Omega=\xi$. Identifying $ns=2Nb=L$, where $L$ is the net spatial extent of our layered $PT$-symmetric system, the  final expression of transmission coefficient $t_{n}=t$ is given by
\beq
t=\frac{e^{-ikL}}{H(k)}
\label{t_exp}
\eeq
where $H(k)=(\xi-i\chi ) U_{N-1}(\Omega)-U_{N-2}(\Omega)$ (Eq. \ref{t_simple}). 
\\
\begin{center}
	{\large \bf Appendix - C: Limiting values of the terms of series expansion of  $ g \chi$ }
\end{center}
The expressions for $A_{1}$ is ,
\beq
A_{1}= N \rho (U_{-} \cos^{2}{\phi} + U_{+} \sin^{2}{\phi}) .
\eeq
Thus, 
\beq
A_{1} b= \frac{L}{2} \rho (U_{-} \cos^{2}{\phi} + U_{+} \sin^{2}{\phi}) .
\eeq
Where we have used $Nb = L/2$. Upon substituting the expressions for $U_{+}$ and $U_{-}$ and using trigonometric identity we arrive at,
\beq
A_{1} b= \frac{L}{2}  (k -\frac{\rho^{2}}{k} \cos{2 \phi} ) .
\eeq
Further substituting $\cos{2 \phi}= \frac{u-k^2}{\rho^2}$ in the above,  we find
\beq
A_{1} b= \frac{k^{2} - q^{2}}{2 k q} (qL) ,
\eeq
where $q= \sqrt{u-k^{2}}$. 
\paragraph{Evaluation of $\lim_{N \to \infty} A_{3} b^{3} $ :} 
From Eq. \ref{a3} we can write, 
\beq
A_{3} b^{3}= - \frac{1}{6} N b^{3} \rho^{3} \left [ 8 N^{2} (U_{+} + U_{-}) \cos{2 \phi} -(U_{+}-U_{-}) \{ 4 N^{2} -1 + (4 N^{2} +1) \cos{4 \phi} \} \right ] .
\eeq
Taking $N^{2}$ out from the parenthesis,  the above equation can be written as , 
\beq
 A_{3} b^{3}= - \frac{1}{6} N^{3} b^{3} \rho^{3} \left [ 8  (U_{+} + U_{-}) \cos{2 \phi} -(U_{+}-U_{-}) \{ 4  -\frac{1}{N^{2}} + (4 +\frac{1}{N^{2}}) \cos{4 \phi} \} \right ].
\eeq
Taking limit $N \rightarrow \infty$ of the above equation,  we get
\beq
\lim_{N \to \infty} A_{3} b^{3}= - \frac{1}{6} N^{3} b^{3} \rho^{3} \left [ 8  (U_{+} + U_{-}) \cos{2 \phi} -(U_{+}-U_{-}) \{ 4   + 4  \cos{4 \phi} \} \right ] .
\eeq
Upon substituting the values of $U_{\pm}$, $\cos{4\phi} = 2 \cos^{2}{2 \phi} -1$, $\cos{2 \phi} = \frac{q^2}{\rho^2}$ and $Nb = L/2$, the above expressions is simplified to ,
\beq
\lim_{N \to \infty} A_{3} b^{3}= - \left ( \frac{ k^{2} -q^{2} }{2 k q}\right ) \frac{(qL)^{3}}{3}.
\eeq
This is the same result given in Eq. \ref{a3b}. 
\paragraph{Evaluation of $\lim_{N \to \infty} A_{5} b^{5}$:} 
From Eq. \ref{a5}, we write
\begin{multline}
A_{5} b^{5}= \frac{N b^{5} \rho ^{5}}{360} [ 2 (U_{+} + U_{-}) (96 N^{4} -5 N^{2} -1) - (U_{+} - U_{-}) \cos{2 \phi} (288 N^{4} -25 N^{2} -8) + \\  2 (U_{+} + U_{-}) \cos{4 \phi} (96 N^{4} + 5 N^{2} +1) - (U_{+} - U_{-}) \cos{6 \phi} (96 N^{4} + 25 N^{2} +8)  ] .
\end{multline}
This can be further written as, 
\begin{multline}
A_{5} b^{5}= \frac{N^{5} b^{5} \rho ^{5}}{360} [ 2 (U_{+} + U_{-}) (96  - \frac{5}{N^{2}} -\frac{1}{N^{4}}) - (U_{+} - U_{-}) \cos{2 \phi} (288 - \frac{25}{N^{2}} -\frac{8}{N^{4}}) + \\  2 (U_{+} + U_{-}) \cos{4 \phi} (96  +  \frac{5}{N^{2}} +\frac{1}{N^{4}}) - (U_{+} - U_{-}) \cos{6 \phi} (96  + \frac{25} {N^{2}} + \frac{8}{N^{4}} )  ] .
\end{multline}
Taking the limit $N \rightarrow \infty$ of the above equation, all terms containing $N$ in denominator will become zero and we get
\beq
\lim_{N \to \infty} A_{5} b^{5}= \frac{L^{5} \rho ^{5}}{2^{5} 360} [ 768 \frac{k}{\rho} \cos^{2}{2 \phi} -192 \frac{\rho}{k} (3 \cos{2 \phi} + \cos{6 \phi}) ]
\eeq
In the above we have already used $Nb = L/2$ and the expressions for $U_{\pm}$. Next we expand $\cos{4 \phi}$ and $\cos{6 \phi}$ in the power of $\cos{2 \phi}$ and substitute $\cos{2 \phi} = \frac{q^{2}}{\rho^2}$ to arrive at
\beq
\lim_{N \to \infty} A_{5} b^{5} = \frac{2}{15} \left ( \frac{k^{2}-q^{2}}{2kq} \right )  ( q L)^{5},
\eeq
which is Eq. \ref{a5b}.

\paragraph{Evaluation of $\lim_{N \to \infty} A_{7} b^{7} $:}
From Eq. \ref{a7} we can write,
\begin{multline}
A_{7} b^{7}= \frac{N b^{7} \rho ^{7}}{15120} [  (U_{+} - U_{-}) (9792 N^{6} -1008 N^{4} -161 N^{2} -34) -\\ 4 (U_{+} + U_{-}) \cos{2 \phi} (4896 N^{6} -168 N^{4} -35 N^{2} -10) + \\ 4 (U_{+} - U_{-}) \cos{4 \phi} (3264 N^{6} -35 N^{2} -16) - \\ 4 (U_{+} + U_{-}) \cos{6 \phi} (1632 N^{6} + 168 N^{4} + 35 N^{2} +10) + \\ (U_{+} - U_{-}) \cos{8 \phi} (3264 N^{6} + 1008 N^{4} + 301 N^{2} +98)   ] .
\end{multline}
The above equation can be further written as ,
\begin{multline}
A_{7} b^{7}= \frac{N^{7} b^{7} \rho ^{7}}{15120} [  (U_{+} - U_{-}) (9792  -\frac{1008}{ N^{2}} - \frac{161} {N^{4}} -\frac{34}{N^{6}}) -\\ 4 (U_{+} + U_{-}) \cos{2 \phi} (4896  -\frac{168} {N^{2}} -\frac{35} {N^{4}} - \frac{10}{N^{6}}) + \\ 4 (U_{+} - U_{-}) \cos{4 \phi} (3264  -\frac{35} {N^{4}} -\frac{16}{N^{6}}) - \\ 4 (U_{+} + U_{-}) \cos{6 \phi} (1632  + \frac{168} {N^{2}} + \frac{35} {N^{4}} +\frac{10}{N^{6}}) + \\ (U_{+} - U_{-}) \cos{8 \phi} (3264  + \frac{1008} {N^{2}} + \frac{301} {N^{4}} +\frac{98}{N^{6}})   ] .
\end{multline}
Taking $N \rightarrow \infty$ limit of the above equation and substituting $Nb=\frac{L}{2}$, we obtain
\begin{multline}
\lim_{N \to \infty} A_{7} b^{7}= \frac{L^{7}  \rho ^{7}}{2^{7} . 15120} [   9792(U_{+} - U_{-})   - 19584 (U_{+} + U_{-}) \cos{2 \phi} + 13056  (U_{+} - U_{-}) \cos{4 \phi} \\ -  6528 (U_{+} + U_{-}) \cos{6 \phi}  + 3264 (U_{+} - U_{-}) \cos{8 \phi}   ] .
\end{multline}
Next we expand $\cos{4 \phi}, \cos{6 \phi}, \cos{8 \phi}$ in powers of $\cos{2 \phi}$ and substitute the expressions for $U_{\pm}$ to show,
\begin{multline}
\lim_{N \to \infty} A_{7} b^{7}= \left (\frac{L}{2} \right )^{7}  \frac{  \rho ^{7}}{15120} \left [  \left(\frac{2\rho}{k} \right ) \{ 9792  + 3264 (8 \cos^{4}{2 \phi}-3)\} - \left (\frac{2k}{\rho} \right ) 261121 \cos^{3}{2 \phi}  \right ] .
\end{multline}
Upon substituting $\cos{2 \phi} = \frac{q^{2}}{\rho^2}$, the above equation simplifies to,
\beq
\lim_{N \to \infty} A_{7} b^{7} =- \frac{17}{315} \left ( \frac{k^{2}-q^{2}}{2kq} \right )  ( q L)^{7},
\eeq
 which is Eq. \ref{a7b}.

\paragraph{Evaluation of $ \lim_{N \to \infty} A_{9} b^{9} $:}
From Eq. \ref{a9} we can write,
\begin{multline}
A_{9} b^{9}= \frac{N b^{9} \rho ^{9}}{453600} [ 4 (U_{+} + U_{-}) (119040 N^{8} - 6120 N^{6} -1029 N^{4} -215 N^{2} -61) - \\ 2 (U_{+} - U_{-}) \cos{2 \phi} (396800 N^{8} -28560 N^{6} -5502 N^{4} - 1045 N^{2} -268) + \\ 40 (U_{+} + U_{-}) \cos{4 \phi} ( 15872 N^{8} -42 N^{4} -15 N^{2} -5) - \\ (U_{+} - U_{-}) \cos{6 \phi} (396800 N^{8} + 28560 N^{6} + 1722 N^{4} - 655 N^{2} -352) + \\ 4 (U_{+} + U_{-}) \cos{8 \phi} (39680 N^{8} + 6120 N^{6} + 1449 N^{4} + 365 N^{2} + 111) - \\ (U_{+} - U_{-}) \cos{10 \phi} (79360 N^{8} + 28560 N^{6} + 9282 N^{4} + 2745 N^{2} + 888)   ].
\end{multline}
Again we write the above equation as follows
\begin{multline}
A_{9} b^{9}= \frac{N^{9} b^{9} \rho ^{9}}{453600} [ 4 (U_{+} + U_{-}) (119040  - \frac{6120} {N^{2}} - \frac{1029}{ N^{4}} -\frac{215} {N^{6}} - \frac{61}{N^{8}}) - \\ 2 (U_{+} - U_{-}) \cos{2 \phi} (396800  - \frac{28560} {N^{2}} -\frac{5502} {N^{4}} - \frac{1045} {N^{6}} -\frac{268}{N^{8}}) + \\ 40 (U_{+} + U_{-}) \cos{4 \phi} ( 15872  - \frac{42} {N^{4}} -\frac{15} {N^{6}} - \frac{5}{N^{8}}) - \\ (U_{+} - U_{-}) \cos{6 \phi} (396800  + \frac{28560} {N^{2}} + \frac{1722} {N^{4}} - \frac{655} {N^{6}} -\frac{352}{N^{8}}) + \\ 4 (U_{+} + U_{-}) \cos{8 \phi} (39680  + \frac{6120} {N^{2}} + \frac{1449} {N^{4}} + \frac{365} {N^{6}} + \frac{111}{N^{8}}) - \\ (U_{+} - U_{-}) \cos{10 \phi} (79360  + \frac{28560} {N^{2}} + \frac{9282} {N^{4}} + \frac{2745} {N^{6}} + \frac{888}{N^{8}}) .
\end{multline}
Taking $N \rightarrow \infty$ limit of the above equation yield,
\begin{multline}
\lim_{N \to \infty}  A_{9} b^{9}= \frac{N^{9} b^{9} \rho ^{9}}{453600} [ 476160(U_{+} + U_{-}) -  793600 (U_{+} - U_{-}) \cos{2 \phi}  +  634880 (U_{+} + U_{-}) \cos{4 \phi}  - \\ 396800 (U_{+} - U_{-}) \cos{6 \phi}    + 158720 (U_{+} + U_{-}) \cos{8 \phi}  - 79360 (U_{+} - U_{-}) \cos{10 \phi} ] .
\end{multline}
We expand $\cos{10 \phi}, \cos{8 \phi}, \cos{6 \phi}, \cos{4 \phi} $ in power of $\cos{2 \phi} = \frac{q^{2}}{\rho^2}$ and substitute the expressions for  $U_{\pm}$ and $Nb =\frac{L}{2}$. It can be shown the above expression finally simplifies to ,
\beq
\lim_{N \to \infty} A_{9} b^{9}=  \frac{31}{2835} \left ( \frac{k^{2}-q^{2}}{2kq} \right )  ( q L)^{9}.
\eeq 
This is Eq. \ref{a9b}. 
\begin{center}
	{\large \bf Appendix - D : Evaluation of $ \left [ \lim_{b \to \infty} \left (\frac{d \tau_{\varepsilon}}{ d\varepsilon} \right ) \right ]_{\varepsilon =1} $ } 
\end{center}
In this appendix, we evaluate the right hand side of Eq. \ref{tau_eps1}. The expression of $P$ is given by Eq. \ref{p_exp}. We express $\frac{dP}{dk}$ , $\frac{dP}{d\varepsilon}$ and $\frac{d^{2}P}{d\varepsilon  dk}$ in terms of the derivatives of $\alpha, Q_{1}$ and $Q_{2}$ as follows,
\begin{multline}
\frac{dP}{df}= \frac{1}{(Q_{1}+ Q_{2} \tan{\alpha})} \left [ Q_{2} \sec^{2}\alpha \frac{d \alpha}{df} + \tan{\alpha} \frac{d Q_{1}}{df} - \frac{d Q_{2}}{df} \right ] - \\ \frac{P}{(Q_{1}+ Q_{2} \tan{\alpha})} \left [ \frac{d Q_{1}}{df} + Q_{2} \sec^{2}{\alpha} \frac{d \alpha}{df} + \tan{\alpha} \frac{d Q_{2}}{df} \right ] .
\label{dpdf}
\end{multline}
Where $f=k, \varepsilon $ .
Also,
\begin{multline}
\frac{d^{2}P}{d \varepsilon dk}=   \frac{1}{(Q_{1}+ Q_{2} \tan{\alpha})}  [ Q_{1} \sec^{2}{\alpha} \frac{d^{2} \alpha}{ d \varepsilon dk} + \sec^{2}{\alpha} \frac{d \alpha}{ dk} \frac{d Q_{1}}{ d \varepsilon} + 2 Q_{1} \sec^{2}{\alpha} \tan{\alpha} \frac{d \alpha}{dk} \frac{d \alpha}{d \varepsilon} \\ + \tan{\alpha} \frac{d^{2} Q_{1}}{ d \varepsilon dk} + \sec^{2}{\alpha} \frac{dQ_{1}}{dk} \frac{d \alpha}{d\varepsilon} -\frac{d^{2} Q_{2}}{d \varepsilon dk} ] \\   - \frac{1}{(Q_{1}+ Q_{2} \tan{\alpha})^{2}} (\frac{d Q_{1}}{d \varepsilon} + \tan{\alpha} \frac{d Q_{2}}{d \varepsilon} + Q_{2} \sec^{2}{\alpha} \frac{d \alpha}{d \varepsilon} ) ( Q_{1} \sec^{2}{\alpha} \frac{d \alpha}{d k} + \tan{\alpha} \frac{d Q_{1}}{dk} - \frac{d Q_{2}}{dk} ) \\ - \frac{P}{(Q_{1}+ Q_{2} \tan{\alpha})} [  \frac{d^{2} Q_{1}}{d \varepsilon dk} + Q_{2} \sec^{2}{\alpha} \frac{d^{2} \alpha}{ d \varepsilon dk} + \sec^{2}{\alpha} \frac{d \alpha}{dk} \frac{d Q_{2}}{d \varepsilon } + \\  2 Q_{2} \sec^{2}{\alpha} \tan{\alpha} \frac{d \alpha}{dk} \frac{d \alpha}{d \varepsilon} + \sec^{2}{\alpha} \frac{d \alpha}{d \varepsilon} \frac{d Q_{2}}{dk} + \tan{\alpha} \frac{d^{2} Q_{2}}{ d \varepsilon dk}] \\ - ( \frac{d Q_{1}}{dk} + Q_{2} \sec^{2}{\alpha} \frac{d \alpha}{dk} + \tan{\alpha} \frac{d Q_{2}}{dk}) [ \frac{1}{(Q_{1}+ Q_{2} \tan{\alpha})} \frac{dP}{d \varepsilon } \\ - \frac{P}{(Q_{1}+ Q_{2} \tan{\alpha})^{2}} \{ \frac{d Q_{1}}{d \varepsilon} + \frac{d Q_{2}}{d \varepsilon} \tan{\alpha} + Q_{2} \sec^{2}{\alpha} \frac{d \alpha}{d \varepsilon }  \}] .
\label{d2pdkde}
\end{multline}
Various derivatives of $\alpha$, $Q_{1}, Q_{2}$ are given by,
\beq
\frac{d \alpha}{dk} =  b k \left[\frac{\left(k^2-u\right) \cos {\phi_{1}}-v \sin {\phi_{1}}}{\rho_{1}^{3}}+\frac{v \epsilon  \sin{\phi_{2}}-\left(k^2-u\right) \cos{\phi_{1}}}{\rho_{2}^{3}}\right].
\label{dalphadk}
\eeq
\beq
\frac{d \alpha}{d \varepsilon }= -\frac{b v^2 \epsilon }{\left(2 \sqrt{2} \rho_{2}^2\right) \sqrt{\rho_{2}^2-\left(k^2-u\right)}} .
\label{dalphadeps}
\eeq
\beq
\frac{d^2 \alpha  }{d\varepsilon dk}=\frac{b k v}{2 \rho _2^7} \left[\sin \phi _2 \left(\left(k^2-u\right)^2-v^2 \epsilon ^2\right)+2 v \epsilon  \left(k^2-u\right) \cos \phi _2\right] .
\label{d2alphadepsdk}
\eeq
\begin{multline}
\frac{dQ_{1}}{dk}=\frac{1}{k^2 \rho _1^5 \rho _2^5} \Big[ \rho _1^5 \left(k^2-\rho _2^2\right) \cos \left(\phi _2\right) \left(u \left(k^2-u\right)-v^2 \epsilon ^2\right)-\rho _2^5 \left(k^2-\rho _1^2\right) \cos \left(\phi _1\right) \left(u \left(k^2-u\right)-v^2\right) \\ -k^2 \rho _2^5 v \left(k^2+\rho _1^2\right) \sin \left(\phi _1\right)-k^3 \left(\rho _1^2-\rho _2^2\right) v^2 \left(\epsilon ^2-1\right) \left(k^2-u\right) \cos \left(\phi _1+\phi _2\right)\\ +k^2 \rho _1^5 v \epsilon  \left(k^2+\rho _2^2\right) \sin \left(\phi _2\right)+k^3 \left(\rho _1^2+\rho _2^2\right) v (\epsilon +1) \sin \left(\phi _1+\phi _2\right) \left(\left(k^2-u\right)^2+v^2 \epsilon \right) \Big] .
\label{dq1dk}
\end{multline}
\begin{multline}
\frac{dQ_{1}}{d \varepsilon }=\frac{v}{2 k \rho _1 \rho _2^5} \Big[\rho _1 v \epsilon  \left(k^2-\rho _2^2\right) \cos \left(\phi _2\right)+k \left(\rho _1^2-\rho _2^2\right) v \epsilon  \cos \left(\phi _1+\phi _2\right) \\ -\left(k^2-u\right) \left(\rho _1 \left(k^2+\rho _2^2\right) \sin \left(\phi _2\right)+k \left(\rho _1^2+\rho _2^2\right) \sin \left(\phi _1+\phi _2\right)\right) \Big ] .
\label{dq1deps}
\end{multline}
\begin{multline}
\frac{d^{2}Q_{1}}{d \varepsilon dk}= \frac{1}{2 k^2 \rho _1^9 \rho _2^{17}} \Big[ -k^2 \rho _1^9 \rho _2^8 v^2 \epsilon  \left(k^2+\rho _2^2\right) \left(k^2-u\right) \cos \left(\phi _2\right) - \\ \rho _1^9 \rho _2^8 v^2 \epsilon  \cos \left(\phi _2\right) \left(-k^2 \rho _2^4-3 k^2 \rho _2^2 \left(k^2-u\right)+5 k^4 \left(k^2-u\right)-\rho _2^6\right) \\ -k^3 \rho _1^4 \rho _2^8 \left(\rho _1^2+\rho _2^2\right) v^2 (\epsilon +1) \left(k^2-u\right) \cos \left(\phi _1+\phi _2\right) \left(\left(k^2-u\right)^2+v^2 \epsilon \right) \\ -4 k^2 \rho _2^8 \rho _1^9 v^3 \epsilon ^2 \left(k^2+\rho _2^2\right) \sin \left(\phi _2\right)-k^3 \rho _2^8 \left(5 \rho _1^6-3 \rho _2^2 \rho _1^4-\rho _2^4 \rho _1^2-\rho _2^6\right) \rho _1^4 v^2 \epsilon  \left(k^2-u\right) \cos \left(\phi _1+\phi _2\right) \\ \rho _2^8 \rho _1^9 v \left(k^2-\rho _2^2\right) +\left(k^2-u\right) \sin \left(\phi _2\right) \left(u \left(k^2-u\right)-v^2 \epsilon ^2\right)+k^2 \rho _2^8 \rho _1^9 v^3 \epsilon ^2 \left(\rho _2^2-k^2\right) \sin \left(\phi _2\right) \\ +2 k^2 \rho _1^9 \rho _2^{12} v \left(k^2+\rho _2^2\right) \sin \left(\phi _2\right)-k^3 \rho _1^4 \rho _2^8 \left(\rho _1^2-\rho _2^2\right) v^3 \epsilon  (\epsilon +1) \sin \left(\phi _1+\phi _2\right) \left(\left(k^2-u\right)^2+v^2 \epsilon \right) \\  -k^3 \rho _1^4 \left(\rho _1^2-\rho _2^2\right) v^3 \left(\epsilon ^2-1\right) \sin \left(\phi _1+\phi _2\right) \left(v^2 \epsilon ^2 \left(k^2-u\right)+\left(k^2-u\right)^3\right)^2 \\ +2 k^3 \rho _1^8 \rho _2^8 \left(\rho _1^2+\rho _2^2\right) v \sin \left(\phi _1+\phi _2\right) \left(\left(k^2-u\right)^2-v^2 \epsilon ^2\right) \Big] .
\label{d2q1depsdk}
\end{multline}
\begin{multline}
\frac{dQ_{2}}{d k }= \frac{1}{k^2 \rho _1^5 \rho _2^5} \Big[ k^2 \rho _2^5 v \left(\rho _1^2-k^2\right) \cos \left(\phi _1\right)+k^2 \rho _1^5 v \epsilon  \left(\rho _2^2-k^2\right) \cos \left(\phi _2\right) \\ - \rho _2^5 \left(k^2+\rho _1^2\right) \sin \left(\phi _1\right) \left(u \left(k^2-u\right)-v^2\right)-k^3 \left(\rho _1^2-\rho _2^2\right) v (\epsilon +1) \cos \left(\phi _1+\phi _2\right) \left(\left(k^2-u\right)^2+v^2 \epsilon \right) \\ \rho _1^5 \left(k^2+\rho _2^2\right) \sin \left(\phi _2\right) \left(u \left(k^2-u\right)-v^2 \epsilon ^2\right)-k^3 \left(\rho _1^2+\rho _2^2\right) v^2 \left(\epsilon ^2-1\right) \left(k^2-u\right) \sin \left(\phi _1+\phi _2\right)  \Big ] .
\label{dq2dk}
\end{multline}
\begin{multline}
\frac{dQ_{2}}{d \varepsilon }= \frac{v}{2 k \rho _1 \rho _2^5}  \Big[ \rho _1 \left(k^2-\rho _2^2\right) \left(k^2-u\right) \cos \left(\phi _2\right)+k \left(\rho _1^2-\rho _2^2\right) \left(k^2-u\right) \cos \left(\phi _1+\phi _2\right) \\ +v \epsilon  \left(\rho _1 \left(k^2+\rho _2^2\right) \sin \left(\phi _2\right)+k \left(\rho _1^2+\rho _2^2\right) \sin \left(\phi _1+\phi _2\right)\right) \Big ] .
\label{dq2deps}
\end{multline}
\begin{multline}
\frac{d^{2}Q_{2}}{d \varepsilon dk }= \frac{1}{2 k^2 \rho _1^9 \rho _2^{17}} \Big [ 4 k^2 \rho _1^9 \rho _2^8 v^3 \epsilon ^2 \left(k^2-\rho _2^2\right) \cos \left(\phi _2\right)+2 k^2 \rho _1^9 \rho _2^{12} v \left(\rho _2^2-k^2\right) \cos \left(\phi _2\right) \\ +k^2 \rho _1^9 \rho _2^8 v^3 \epsilon ^2 \left(k^2+\rho _2^2\right) \cos \left(\phi _2\right)-\rho _1^9 \rho _2^8 v \left(k^2+\rho _2^2\right) \left(k^2-u\right) \cos \left(\phi _2\right) \left(u \left(k^2-u\right)-v^2 \epsilon ^2\right) \\ -2 k^3 \rho _1^8 \rho _2^8 \left(\rho _1^2-\rho _2^2\right) v \cos \left(\phi _1+\phi _2\right) \left(\left(k^2-u\right)^2-v^2 \epsilon ^2\right) \\ +k^3 \rho _1^4 \rho _2^8 \left(\rho _1^2+\rho _2^2\right) v^3 \epsilon  (\epsilon +1) \cos \left(\phi _1+\phi _2\right) \left(\left(k^2-u\right)^2+v^2 \epsilon \right) \\ +k^3 \rho _1^4 \left(\rho _1^2+\rho _2^2\right) v^3 \left(\epsilon ^2-1\right) \cos \left(\phi _1+\phi _2\right) \left(v^2 \epsilon ^2 \left(k^2-u\right)+\left(k^2-u\right)^3\right)^2 \\ \rho _1^9 \rho _2^8 \left(-v^2\right) \epsilon  \sin \left(\phi _2\right) \left(-k^2 \rho _2^4+3 k^2 \rho _2^2 \left(k^2-u\right)+5 k^4 \left(k^2-u\right)+\rho _2^6\right) \\ +k^2 \rho _1^9 \rho _2^8 v^2 \epsilon  \left(\rho _2^2-k^2\right) \left(k^2-u\right) \sin \left(\phi _2\right)\\ -k^3 \rho _1^4 \rho _2^8 \left(\rho _1^2-\rho _2^2\right) v^2 (\epsilon +1) \left(k^2-u\right) \sin \left(\phi _1+\phi _2\right) \left(\left(k^2-u\right)^2+v^2 \epsilon \right) \\ -k^3 \rho _1^4 \rho _2^8 \left(5 \rho _1^6+3 \rho _2^2 \rho _1^4-\rho _2^4 \rho _1^2+\rho _2^6\right) v^2 \epsilon  \left(k^2-u\right) \sin \left(\phi _1+\phi _2\right) \Big ].
\label{d2q2depsdk}
\end{multline}
The right hand side of Eq. \ref{tau_eps1} is to be evaluated at $\varepsilon =1$. Therefore, we evaluate all the above derivatives of $\alpha$, $Q_{1}$ and $Q_{2}$ at $\varepsilon =1$. When $\varepsilon =1$, we also have $\rho_{2}= \rho_{1}$, $\phi_{2}= \phi_{1}$ and $\alpha=0$. We simplify the Eqs. \ref{dalphadk}, \ref{dalphadeps}, \ref{d2alphadepsdk}, \ref{dq1dk}, \ref{dq1deps}, \ref{d2q1depsdk}, \ref{dq2dk}, \ref{dq2deps} and Eq. \ref{d2q2depsdk} at $\varepsilon =1$ to obtain the following results,
\beq
\left[ \frac{d \alpha}{dk} \right]_{\varepsilon =1} =0 .
\label{dalphadke1}
\eeq
\beq
\left[ \frac{d Q_{1}}{dk} \right]_{\varepsilon =1}=\frac{4 k v^2}{\rho _1^6} .
\label{dq1dke1}
\eeq
\beq
 \left[ \frac{d Q_{2}}{d k } \right]_{\varepsilon =1} =\frac{2 \left(k^2+\rho _1^2\right) \sin \left(\phi _1\right) \left(u \left(k^2-u\right)-v^2\right)}{k^2 \rho _1^5}+\frac{(2 v) \left(\rho _1^2-k^2\right) \cos \left(\phi _1\right)}{\rho _1^5} .
\label{dq2dke1}
\eeq
\beq
\left[ \frac{d \alpha}{d \varepsilon } \right]_{\varepsilon =1} = -\frac{b v^2}{2 \sqrt{2} \rho _1^2} \left [ \frac{1}{\sqrt{\rho _1^2-\left(k^2-u\right)}}\right ] .
\label{dalphadepse1}
\eeq
\beq
\left[ \frac{d Q_{1}}{d \varepsilon } \right]_{\varepsilon =1}= \left (\frac{v}{2 k \rho _1^5} \right ) \left [  v \left(k^2-\rho _1^2\right) \cos \left(\phi _1\right)-\left(k^2-u\right) \sin \left(\phi _1\right) \left(k^2+4 k \rho _1 \cos \left(\phi _1\right)+\rho _1^2\right)\right ] .
\label{dq1depse1}
\eeq
\beq
\left[ \frac{d Q_{2}}{d \varepsilon } \right]_{\varepsilon =1} =\left ( \frac{v}{2 k \rho _1^5} \right ) \left[ \left(k^2-\rho _1^2\right) \left(k^2-u\right) \cos \left(\phi _1\right)+v \left(k^2+\rho _1^2\right) \sin \left(\phi _1\right)+\frac{2 k v^2}{\rho _1} \right] .
\label{dq2depse1}
\eeq
\beq
\left[ \frac{d^{2} \alpha}{d \varepsilon dk} \right]_{\varepsilon =1} = \frac{b k v}{2 \rho _1^7} \left [ \sin \left(\phi _1\right) \left(\left(k^2-u\right)^2-v^2\right)+2 v \left(k^2-u\right) \cos \left(\phi _1\right) \right ] .
\label{d2alphadepsdke1}
\eeq
\begin{multline}
\left[ \frac{d^{2} Q_{1}}{d \varepsilon dk} \right]_{\varepsilon =1} = \Big [ \rho _1^3 v \cos \left(\phi _1\right) \left(-6 k^6+2 k^4 \left(\rho _1^2+3 u\right)+k^2 \rho _1^2 \left(\rho _1^2-2 u\right)+\rho _1^6\right) \\  -4 k^3 \rho _1^4 v \left(k^2-u\right) \cos \left(2 \phi _1\right)  +4 k^3 \rho _1^3 \rho _1 \sin \left(2 \phi _1\right) \left(\left(k^2-u\right)^2-v^2\right) \\ + \rho _1^3 \sin \left(\phi _1\right) \Big\{2 k^2 \rho _1^4 \left(k^2+\rho _1^2\right)-\rho _1^2 \left(v^2 \left(2 k^2+u\right)+u \left(k^2-u\right)^2\right)+ \\ k^2 \left(v^2 \left(u-6 k^2\right)+u \left(k^2-u\right)^2\right)\Big\} \Big ] .
\label{d2q1depsdke1}
\end{multline}
\begin{multline}
\left[ \frac{d^{2} Q_{2}}{d \varepsilon dk} \right]_{\varepsilon =1} = \frac{v}{2 k^2 \rho _1^{14}} \Big [ 4 k^3 v^2 \left(k^2-u\right) \left(\left(k^2-u\right)^2+2 \rho _1^4+v^2\right) \\ + \rho _1^5 v \sin \left(\phi _1\right) \left(-6 k^6+k^4 \left(6 u-2 \rho _1^2\right)+k^2 \rho _1^2 \left(\rho _1^2+2 u\right)-\rho _1^6\right) \\ - \rho _1^5 \cos \left(\phi _1\right) \Big\{2 k^2 \rho _1^4 \left(k^2-\rho _1^2\right)+\rho _1^2 \left(v^2 \left(2 k^2+u\right)+u \left(k^2-u\right)^2\right)+ \\ k^2 \left(v^2 \left(u-6 k^2\right)+u \left(k^2-u\right)^2\right)\Big\}   \Big ] .
\label{d2q2depsdke1}
\end{multline}
For $\varepsilon =1$ we have the following simplifications,
\beq
Q_{1}(\varepsilon =1)= 4 \sin^{2}{\phi_{1}}, \ \ Q_{2}(\varepsilon =1)= 2 J_{1}^{-} \sin{\phi_{1}} ,
\label{q1q2_e1}
\eeq
and,
\beq
P (\varepsilon =1)= -\frac{1}{2} J_{1}^{-} \csc{\phi_{1}}
\label{p_e1}
\eeq
From the results of Eqs. \ref{dalphadke1},\ref{dalphadepse1}, \ref{d2alphadepsdke1}, \ref{dq1dke1}, \ref{dq1depse1}, \ref{d2q1depsdke1}, \ref{dq2dke1}, \ref{dq2depse1}, \ref{d2q2depsdke1} and Eqs. \ref{q1q2_e1}, \ref{p_e1} we can simplify $(\frac{dP}{dk})_{\varepsilon =1}, (\frac{dP}{d \varepsilon })_{\varepsilon =1}$ and $(\frac{d^{2}P}{d \varepsilon dk})_{\varepsilon =1}$ and can evaluate the right hand side of Eq. \ref{tau_eps1}. After a lengthy algebra, it can be shown that ,
\beq
\left [ \frac{1}{(1+ P^{2})}  \frac{d^{2}P}{d \varepsilon dk} - \frac{2P}{(1+ P^{2})^{2}} \frac{dP}{dk} \frac{dP}{d \varepsilon}\right ]_{\varepsilon =1} = K_{0} + K_{1} b.
\eeq 
Where the expressions for $K_{1}$ and $K_{0}$ are given by,
\beq
K_{1}= \frac{kv}{2 \rho_{1}^{7}} \left [ \{(k^{2}-u)^{2} -v^{2}\} \sin{\phi_{1}} + 2 (k^{2}-u) v \cos{\phi_{1}} \right]
\label{k1_exp}.
\eeq
The expression for $K_{0}$ is lengthy and we expressed through the use of symbols $C_{1}, C_{2}, C_{3}, C_{4}, C_{5}$ and $C_{6}$ as given below,
\beq
K_{0} =\frac{v \csc ^2\left(\phi _1\right)}{2 k^3 \rho _1^{13} \left[\left( J_{1}^{-}\right){}^2 \csc ^2\left(\phi _1\right)+4\right]{}^2} (C_{1}+ C_{2}+ C_{3}+ C_{4}+ C_{5}- C_{6}).
\label{k0_exp}
\eeq
Various $C_{i}$'s appearing in the above equation are given by,
\begin{multline}
C_{1} =2 k^3 v \cot \left(\phi _1\right) \left(k^4-2 k^2 u+u^2+v^2\right) (-\left( J_{1}^{-}\right)^2-2 \cos \left(2 \phi _1\right)+2) \\ \times \cot \left(\phi _1\right) \left(\left(k^2-\rho _1^2\right) \left(k^2-u\right) \cot \left(\phi _1\right)+v \left(k^2+\rho _1^2\right)+4 k \rho _1 v \cos \left(\phi _1\right)\right) .
\end{multline}
\begin{multline}
C_{2} = \left [8 \left( J_{1}^{-}\right) k^3 v \cot \left(\phi _1\right) \left(k^4-2 k^2 u+u^2+v^2\right) \right ] \\ \times \left [ 4 k \rho _1 \left(k^2-u\right) \cos \left(\phi _1\right)+\left(k^2+\rho _1^2\right) \left(k^2-u\right)-v \left(k^2-\rho _1^2\right) \cot \left(\phi _1\right)  \right ] .
\end{multline}
\begin{multline}
C_{3} = \rho _1^2 \csc ^2\left(\phi _1\right) \left ( 2-\frac{1}{2} \left( J_{1}^{-}\right)^2 \csc ^2\left(\phi _1\right) \right ) \\ \times \left(k^2+\rho _1^2\right) \sin \left(\phi _1\right) \left(u \left(k^2-u\right)-v^2\right)+k^2 v \left(\rho _1^2-k^2\right) \cos \left(\phi _1\right) \\ \times \rho _1 v \left(k^2-\rho _1^2\right) \cos \left(\phi _1\right)-\rho _1 \left(k^2-u\right) \sin \left(\phi _1\right) \left(k^2+4 k \rho _1 \cos \left(\phi _1\right)+\rho _1^2\right).
\end{multline}
\begin{multline}
C_{4} = 2  J_{1}^{-} \rho _1^2 \csc ^3\left(\phi _1\right) \left(\left(k^2+\rho _1^2\right) \sin \left(\phi _1\right) \left(u \left(k^2-u\right)-v^2\right)+k^2 v \left(\rho _1^2-k^2\right) \cos \left(\phi _1\right)\right) \\ \times \rho _1 \left(k^2-\rho _1^2\right) \left(k^2-u\right) \cos \left(\phi _1\right)+\rho _1 v \sin \left(\phi _1\right) \left(k^2+4 k \rho _1 \cos \left(\phi _1\right)+\rho _1^2\right).
\end{multline}
\begin{multline}
C_{5} = 2  J_{1}^{-} k \rho _1 \csc \left(\phi _1\right) \left ( \frac{1}{4} \left (J_{1}^{-} \right)^2 \csc ^2\left(\phi _1\right)+1 \right ) \\ \times \Big [ \rho _1^3 v \cos \left(\phi _1\right)   \left(-6 k^6+2 k^4 \left(\rho _1^2+3 u\right)  +k^2 \left(\rho _1^4-2 \rho _1^2 u\right)+\rho _1^6 \right)  \\ -4 k^3 v \left(k^2-u\right) \cos \left(2 \phi _1\right) \left(k^4-2 k^2 u+u^2+v^2\right) \\ \rho _1^3 \sin \left(\phi _1\right) \Big \{ k^6 u-k^4 \left(-2 \rho _1^4+2 u^2+\rho _1^2 u+6 v^2\right)-\rho _1^2 u \left(u^2+v^2\right) \\ +k^2 \left(2 \rho _1^6+u^3+2 \rho _1^2 u^2+u v^2-2 \rho _1^2 v^2\right)+8 k^3 \rho _1 \cos \left(\phi _1\right) \left(k^4-2 k^2 u+u^2-v^2\right) \Big \}  \Big ]
\end{multline}
\begin{multline}
C_{6} = 4 k \rho _1 \left(\frac{1}{4} \left (J_{1}^{-} \right)^2 \csc ^2\left(\phi _1\right)+1\right) \Big [ -\rho _1^3 \cos \left(\phi _1\right) \\ \times  \Big \{k^6 u+k^4 \left(2 \rho _1^4-2 u^2+\rho _1^2 u-6 v^2\right)+k^2 \left(-2 \rho _1^6+u^3-2 \rho _1^2 u^2+u v^2+2 \rho _1^2 v^2\right)+\rho _1^2 u \left(u^2+v^2\right)  \Big \} \\ + v \Big \{ 4 k^3 v \cos \left(2 \phi _1\right) \left(k^4-2 k^2 u+u^2+v^2\right) \\ -\rho _1^3 \sin \left(\phi _1\right) \left(6 k^6+k^4 \left(2 \rho _1^2-6 u\right)-k^2 \left(\rho _1^4+2 \rho _1^2 u\right)+16 k^3 \rho _1 \left(k^2-u\right) \cos \left(\phi _1\right)+\rho _1^6\right) \Big \}   \Big ].
\end{multline}
$K_{0}$ and $K_{1}$ are independent of the thickness `$2b$' .


\begin{thebibliography}{1}
\bibitem{ben4} C. M. Bender and S. Boettcher, {\em Phys. Rev. Lett.} {\bf 80}, 5243 (1998).

\bibitem{mos} A. Mostafazadeh, {\em Int. J. Geom. Meth. Mod. Phys.} {\bf 7}, 1191(2010) and references therein.

\bibitem{benr} C.M. Bender, {\em Rep. Progr. Phys.} {\bf 70} (2007) 947 and references therein.

\bibitem {nh1} J. Christensen, M. Willatzen, V. R. Velasco, and M.H. Lu, {\em Phys. Rev. Lett.  } {\bf 116} , 207601 (2016)  

\bibitem {nh2} J. Schindler, A. Li, M. C. Zheng, F. M. Ellis, and T. Kottos , { \em Phys. Rev. A} {\bf 84}, 040101 (2011).

\bibitem {nh3} Y. Xu, S. Wang, and L.-M. Duan, { \em Phys. Rev. Lett.} { \bf 118}, 045701 (2017). 

\bibitem {nh4} J. Xu, Y. Du, W. Huang, and D. Zhang,{ \em Opt. Exp.} {\bf 25}, 15786 (2017). 
\bibitem{nh41}  H Rawal, B. P. Mandal, Nucl. Phys. B946, 114699 (2019).
\bibitem{nh42} A Khare, B. P. Mandal, Phys. Lett A 272, 53 (2000).

\bibitem {nh5} M. Hasan,  B. P. Mandal, {\em Ann. of Phys.} , 396, 371 (2018).

\bibitem {nh6} T. E. Lee, {\em Phys. Rev. Lett.} { \bf 116}, 133903 (2016). 

\bibitem {nh7} A. Ghatak, M. Hasan, B.P. Mandal, {\bf Phys. Lett. A} {\bf 379}, 1326 (2015). 

\bibitem{opt1} Z. H. Musslimani, K. G. Makris, R. El-Ganainy, and D. N. Christodoulides, {\em Phys. Rev. Lett.} {\bf 100}, 030402 (2008).

\bibitem{opt2} C. E. Ruter, K. G. Makris, R. El-Ganainy, D. N. Christodoulides, M. Segev, D. Kip, {\em Nature Phys.} {\bf 6} 192, (2010); 

\bibitem{opt3}  R. El-Ganainy, K. G. Makris, D. N. Christodoulides and Z. H. Musslimani, {\em Opt. Lett.} {\bf 32}, 2632 (2007).

\bibitem {bloch} Y. Xu, W. S. Fegadolli, L. Gan, M. Lu, X. Liu, Z. Li, A. Scherer, and Y. Chen, {\em  Nat. Commun. }, {\bf 7}, 11319 (2016). 

\bibitem{eqv1} A. Guo et al, {\em Phys. Rev. Lett.} {\bf 103}, 093902 (2009).

\bibitem {bittner} Bittner S, et. al.  {\em Phys. Rev. Lett.}  {\bf 108 } , 024101 (2012).

\bibitem {kotto} Kottos T ,  {\em Nature Physics} { \bf 6} , 166 (2010).

\bibitem {nordheim1928} L. Nordheim, proc. R. Soc. A , 119, 173 (1928).

\bibitem {gurney1928} R. W. Gurney, E.U. Condon, Nature, 122,439 (1928). 

\bibitem {condon} E.U. Condon, Rev. Mod. Phys. 3, 43 (1931).
 
\bibitem {wigner_1955} E.P. Wigner, Phys. Rev. 98, 145 (1955).

\bibitem {david_bohm_1951} D. Bohm, Quantum Theory, Prentice-Hall, New York (1951).

\bibitem {hartman_paper} T. E. Hartman, J. App. Phys. 33, 3427 (1962).

\bibitem {fletcher} J. R. Fletcher, J. Phys. C, 18, L55 (1985).

\bibitem {hg_winful} H. G. Winful, Physics Reports, 436 ,1-69 (2006) and references therein. 

\bibitem {generalized_hartman} V. S. Olkhovsky1, E. Recami and G. Salesi, Euro. Phys. Lett. 57, 879 (2002).

\bibitem {esposito_multi_barrier} S. Esposito,  Phy. Rev. E 67, 016609 (2003)

\bibitem {questions_ghf1} S. Kudaka, S. Matsumoto, Phys. Lett. A, 375, 3259 (2011).

\bibitem {questions_ghf2} V. Milanovic, J. Ranovanovic, Phys. Lett. A, 376, 16, 1401 (2012).

\bibitem {questions_ghf3} S. Kudaka, S. Matsumoto, Phys. Lett. A, 376, 1403 (2012).

\bibitem {sl_prl} A.M. Steinberg, P.G. Kwiat, R.Y. Chiao, Phys. Rev. Lett., 71, 708 (1993).

\bibitem {nimtz} G. Nimtz H. Spieker, H.M. Brodowsky, Phys. Lett. A 222, 125 (1996).

\bibitem {ph} Ph. Balcou and L. Dutriaux Phys. Lett. A,78, 851 (1997).

\bibitem {ragni} L. Ragni, Phys. Rev. E, 79, 046609 (2009).

\bibitem {sattari} F. Sattari and E. Faizabadi AIP Advances 2, 12123 (2012).

\bibitem {longhi1} S. Longhi, M. Marano, P. Laporta, and M. Belmonte Phys. Rev. E, 64, 055602 (2001).

\bibitem {olindo} C. Olindo, M.A. Sagioro, F.M. Matinaga, A. Delgado, C.H. Monken, S. Padua , Optics Comm., 272, 161 (2007).

\bibitem {barbero2000} A.P. L. Barbero, H.E. Hernandez-Figueroa, E. Recami, Phys. Rev. E, 62, 6, 8628 (2000).  

\bibitem {recami2000} E. Recami, F. Fontana, R. Garavaglia, Int. J. Mod. Phys. A, 15, 2793 (2000).

\bibitem {longhi2} S. Longhi, P. Laporta, M. Belmonte, E. Recami, Phys. Rev. E, 65, 046610 (2002).

\bibitem {reshaping} V. S. Olkhovsky,   E.  Recami , Phys. Rep. 214, 6 ,339 (1992), V. S. Olkhovsky,   E.  Recami, F. Raciti, A. K. Zaichenko,  Journal de Physique I, 5,1351 (1995), G. Privitera , G. Salesi , V. S. Olkhovsky ,E. Recami , LANL Archives \# cond-mat/9802126. 

\bibitem {complex_barrier_tunneling} F. Raciti, G. Salesi,  Journal de Physique I, EDP Sciences, 4 (12), 1783 (1994).

\bibitem {tt_sfqm_1} M. Hasan, B.P. Mandal, {\em Phys. Lett. A} {\bf 382} 248 (2018)

\bibitem {tt_sfqm_2} M. Hasan, B.P. Mandal, {\em Eur. Phys. J. Plus} {\bf 135} 248 (2020)

\bibitem {hartman_layered} M. Hasan, B. P. Mandal, { \em Eur.  Phys. J. Plus}   { \bf 135} 84 (2020)

\bibitem {dutta_roy_book} {\it Elements of Quantum Mechanics}, B. Dutta Roy, {\it New Age Science Ltd.} (2009).

\bibitem {griffith_periodic} D. J. Griffiths and C. A. Steinkea, {\em  Am. J. Phys.} 69 (2),  137,(2001).
\end{thebibliography}
\end{document}